\documentclass[aps,pra,twocolumn,showpacs,superscriptaddress,longbibliography]{revtex4-1}

\usepackage{graphicx,amsmath,verbatim,amsfonts}
\usepackage{textcomp,bbm}

\newcommand{\deff}{d_\mathrm{eff}}
\newcommand{\Qeff}{Q_\mathrm{eff}}

\newcommand{\F}{\mathcal{F}}
\newcommand{\um}{\,\textmu m}

\begin{document}
\title{A cavity-enhanced single photon source based on the silicon vacancy center in diamond}

\author{Julia Benedikter}
\author{Hanno Kaupp}
\author{Thomas H\"ummer}
\affiliation{Fakult{\"a}t f{\"u}r Physik, Ludwig-Maximilians-Universit{\"a}t, Schellingstra{\ss}e~4, 80799~M{\"u}nchen, Germany}
\affiliation{Max-Planck-Institut f{\"u}r Quantenoptik,  Hans-Kopfermann-Str.~1, 85748~Garching, Germany}
\author{Yuejiang Liang}
\affiliation{Institut f{\"u}r Organische Chemie, Universit{\"a}t W{\"u}rzburg, Am Hubland, 97074 W{\"u}rzburg, Germany}
\author{Alexander Bommer}
\author{Christoph Becher}
\affiliation{Universit{\"a}t des Saarlandes, Fachrichtung 7.2 (Experimentalphysik), Campus E 2.6, 66123~Saarbr{\"u}cken, Germany}
\author{Anke Krueger}
\affiliation{Institut f{\"u}r Organische Chemie, Universit{\"a}t W{\"u}rzburg, Am Hubland, 97074 W{\"u}rzburg, Germany}
\author{Jason M. Smith}
\affiliation{Department of Materials, University of Oxford, 16 Parks Road, Oxford OX1 3PH, United Kingdom}
\author{Theodor W. H{\"a}nsch}
\affiliation{Fakult{\"a}t f{\"u}r Physik, Ludwig-Maximilians-Universit{\"a}t, Schellingstra{\ss}e~4, 80799~M{\"u}nchen, Germany}
\affiliation{Max-Planck-Institut f{\"u}r Quantenoptik,  Hans-Kopfermann-Str.~1, 85748~Garching, Germany}
\author{David Hunger}
\email[To whom correspondence should be addressed. E-mail: ]{david.hunger@kit.edu}
\affiliation{Fakult{\"a}t f{\"u}r Physik, Ludwig-Maximilians-Universit{\"a}t, Schellingstra{\ss}e~4, 80799~M{\"u}nchen, Germany}
\affiliation{Max-Planck-Institut f{\"u}r Quantenoptik,  Hans-Kopfermann-Str.~1, 85748~Garching, Germany}
\affiliation{Physikalisches Institut, Karlsruher Institut f{\"u}r Technologie, Wolfgang-Gaede-Str.1, 76131 Karlsruhe, Germany}

\date{\today}

\begin{abstract}
	Single photon sources are an integral part of various quantum technologies, and solid state quantum emitters at room temperature appear as a promising implementation. We couple the fluorescence of individual silicon vacancy centers in nanodiamonds to a tunable optical microcavity to demonstrate a single photon source with high efficiency, increased emission rate, and improved spectral purity compared to the intrinsic emitter properties. We use a fiber-based microcavity with a mode volume as small as $3.4~\lambda^3$ and a quality factor of $1.9\times 10^4$ and observe an effective Purcell factor of up to 9.2. We furthermore study modifications of the internal rate dynamics and propose a rate model that closely agrees with the measurements. We observe lifetime changes of up to 31\%, limited by the finite quantum efficiency of the emitters studied here. With improved materials, our achieved parameters predict single photon rates beyond 1~GHz.
\end{abstract}



\maketitle

\section{Introduction}

Single photon sources are a fundamental component of the toolbox for quantum information technologies that promise transformational advances in the communication and processing of information \cite{Gisin02,OBrien09}. There is thus large interest in developing scalable sources that fulfill the requirements of high purity (emitting exactly one photon at a time), high efficiency (obtaining a photon in a collectable mode), high brightness (large maximal excitation rate), and high spectral purity (a narrow, ideally Fourier-transform limited spectrum). Solid-state-based quantum emitters \cite{Shields07,Aharonovich16} have evidenced outstanding properties in this respect.
To achieve high collection efficiencies and large emission rates, coupling the emitter to photonic structures \cite{Schroeder16} such as optical resonators \cite{Michler00,Strauf07,Englund10,Gazzano13,Albrecht13,Kaupp16}, waveguides \cite{Laucht12,Liebermeister14}, or antennas \cite{Chu14,Li15} is desired. This has been demonstrated for various systems like quantum dots \cite{Michler00,Strauf07,Gazzano13,Schlehahn16}, molecules \cite{Steiner07,Toninelli10}, carbon nanotubes \cite{Miura14,Pyatkov16,Jeantet16}, nitrogen vacancy (NV) \cite{Faraon12,Albrecht13,Johnson15,Kaupp16}, and silicon vacancy (SiV) \cite{Lee12,Riedrich12,Riedrich14,Sipahigil16} centers in diamond.
While cryogenic experiments already come close to ideal single photon sources, it remains a challenge to achieve high efficiency and spectral purity under ambient conditions.

Here, we demonstrate an approach to achieve high efficiency, brightness, and spectral purity for a room-temperature source by coupling the emission of single SiV centers to a high quality factor microcavity.
Coupling quantum emitters to optical microcavities \cite{Vahala03} increases the spontaneous emission rate by the Purcell factor $ C = \frac{3(\lambda/n)^3}{4\pi^2}\frac{\Qeff}{V_m}\zeta\left|\frac{\vec{\mu}\vec{E}}{\mu E_0}\right|^2 $, where $ \lambda $ is the wavelength, $ n $ the refractive index, $ V_m $ the mode volume, $\vec{\mu}$ the dipole matrix element, $\vec{E}$ the electric field vector at the location of the dipole, $E_0$ the maximal field in the cavity, and $ \zeta $ the branching ratio into the zero phonon line (ZPL). The effective quality factor $ \Qeff =(Q_c^{-1}+Q_\mathrm{em}^{-1})^{-1} $ includes the cavity quality factor $ Q_c $ and the emitter quality factor $ Q_\mathrm{em} $ obtained from the emitter linewidth \cite{Auffeves10,Meldrum10}, motivating the choice of narrow emitters. In this respect, the SiV \cite{Gali13} is particularly promising, due to its narrow ZPL that carries about 80\% of the oscillator strength ($ \zeta = 0.8$), and an excited state lifetime of $\sim 1$~ns, favoring bright single photon emission \cite{Wang06,Neu11,Neu12,Rogers14}.
The emission is coupled to a well-collectable cavity mode with an efficiency $\beta=C/(C+1)$, which can be near-unity for large $C$. Furthermore, in the bad emitter regime, where the spectral width of the fluorescence is broader than the cavity mode, the spectral emission is determined primarily by the properties of the optical resonator rather than the electronic emitter. This is attractive because it offers potential for exquisite control over photon emission even at ambient conditions.

\section{Experimental methods}

\begin{figure}
	\centering
	\includegraphics[width=\columnwidth]{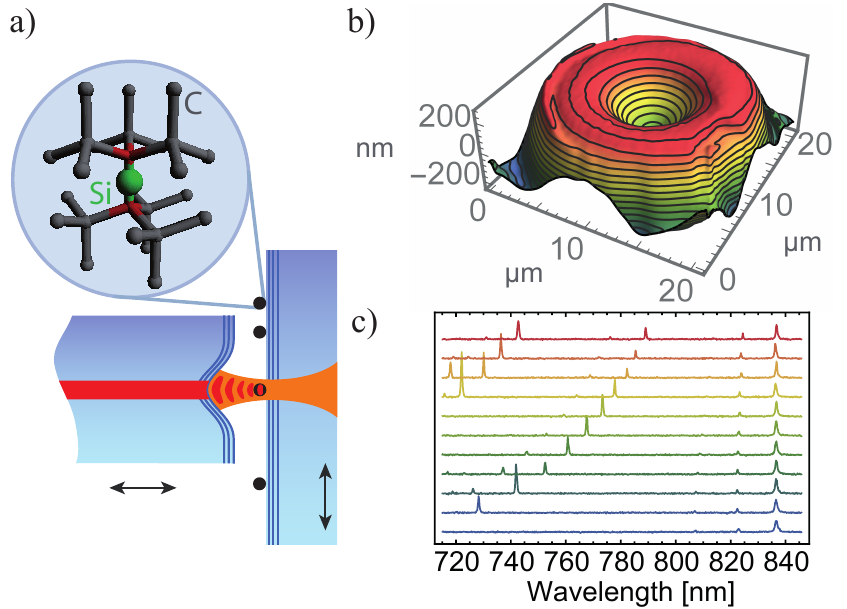}
	\caption{\label{fig:setup} (a) Schematic of cavity and SiV center in diamond lattice (vertical axis is in [111] direction). (b) Interferometric measurement of fiber profile. (c) Cavity transmission spectra for different cavity lengths starting at $ d = 5 \lambda/2 $ (lowest curve) in steps of one free spectral range. Different resonance heights are due to source spectrum and finite spectrometer resolution.  
	}
\end{figure}

We use a fiber-based Fabry-Perot microcavity \cite{Hunger10b} which combines a small mode volume and a high quality factor with full tunability and an open access design, allowing the investigation of different emitters with one and the same cavity.
The cavity consists of a macroscopic planar mirror and a micromirror at the endfacet of a single-mode optical fiber (Fig.~\ref{fig:setup}a). The fiber is shaped by CO$_2$ laser machining \cite{Hunger12} and has a conical tip with a remaining plateau of about $18~\mu$m diameter to allow for the shortest mirror separations \cite{Kaupp16}. In its center, we produce a concave profile with a radius of curvature of $26~\mu$m, (see Fig.~\ref{fig:setup}b). 
The fiber and the planar mirror have different dielectric coatings to optimize excitation through the large mirror and up to 90\% out-coupling of fluorescence to the fiber (see appendix \ref{sec:cavity} for details).
The fiber is mounted on a shear piezo stack, which allows us to accurately tune the cavity length.

We measure the cavity finesse $\F$ at 780~nm and obtain a value which is consistent with $\F=3750$ at 740~nm according to the coating simulation. To calibrate the optical cavity length, we record broad-band cavity transmission spectra with a supercontinuum laser and evaluate the separation of subsequent cavity resonances, see Fig.~\ref{fig:setup}c. We find that the smallest accessible effective cavity length is $\deff=5\lambda/2$, corresponding to the longitudinal mode order $q = 5$, limited by the profile depth (300~nm) and the field penetration into the coating (1160~nm at 740~nm). At this separation, we obtain a cavity quality factor of $ Q_c = q \F = 1.9 \times 10^4$. From scanning-cavity microscopy measurements and calculations \cite{Benedikter15}, we infer the mode waist to be $ w_0 = 1.0~\mu$m, resulting in a mode volume $ V_m = (\pi w_0^2 \deff)/4 = 3.4 \lambda^3 $. Together, these cavity parameters yield an ideal Purcell factor of $C_0=3\lambda^3/(4\pi^2)Q_c/V_m=425$.

\begin{figure}
	\centering
	\includegraphics{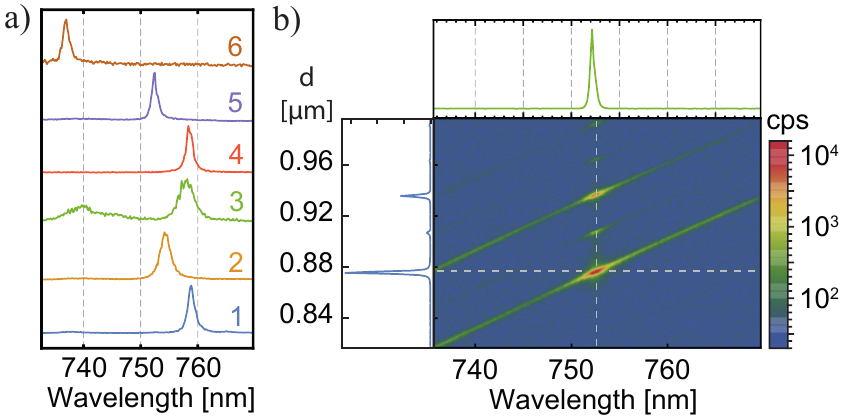}
	\caption{\label{fig:spectra} (a) Free space fluorescence spectra of single SiV centers studied for cavity coupling. (b) Cavity fluorescence spectra of ND5 for different geometric cavity lengths (air gap), logarithmic color scale. Top: Linear scale spectrum for $d=0.875~$nm where maximal enhancement occurs. Left: Linear scale plot of count rate for different lengths at emission wavelength $\lambda=752.4~$nm.
	}
\end{figure}

We study nanodiamonds of about 100~nm in size produced by bead-assisted sonic disintegration of a polycrystalline chemical vapor deposition film \cite{Neu11b}. The nanodiamonds are spin-coated onto the planar mirror and are first investigated by confocal microscopy with an excitation wavelength of 690~nm. The crystals typically contain ensembles of SiV centers featuring a broad (7~nm) ZPL at the nominal wavelength of 738~nm. In some crystals, we also observe narrow (down to 1~nm) emission lines that are spectrally shifted and which show pronounced photon antibunching. We attribute these lines to emission from single SiV centers that are subject to local perturbations such as strain in the nanodiamonds \cite{Neu11b}. It is desirable to study emitters with maximal radiative quantum yield, fluorescence stability, and optimal dipole orientation, which we favor by selecting emitters with high brightness. The spectra of the single SiV centers studied here are shown in Fig.~\ref{fig:spectra}a. The central wavelengths range from 737~nm (the nominal emission wavelength) to 759~nm with spectral widths from 1.1~nm to 3.0~nm (see appendix \ref{sec:data}).

We have developed a setup which combines a confocal microscope and a tunable microcavity side-by-side sharing a single nanopositioner. Calibrating the displacement between the confocal focus and the cavity enables the characterization of the same emitters both in free space and inside the cavity. 
Nanodiamonds pre-characterized confocally can be easily found in the cavity via the Rayleigh scattering and absorption loss they introduce. Therefore, we perform scanning cavity microscopy \cite{Mader15} and measure the cavity transmission of a supercontinuum laser filtered to a 33~nm spectral band around 747~nm. 
On such transmission maps (see appendix \ref{sec:scans}) the nanodiamonds appear as dark spots and can be directly related to the confocal fluorescence maps.

To achieve resonant coupling conditions with the SiV emission, we stepwise tune the cavity length to shift a cavity resonance across the emission spectrum and record the fluorescence spectra on a grating spectrometer. On resonance, we observe enhanced emission into a single cavity resonance. Figure \ref{fig:spectra}b shows fluorescence spectra for varying mirror separation for the $q=5$ mode order in a logarithmic color scale to make the high signal-to-background level visible. Emission away from the cavity resonance is suppressed, 
and we detect only dark counts (blue color). The cavity resonance appears broadened due to the finite spectrometer resolution and some length jitter within the acquisition time of 1~s. The actual FWHM line width is 43~pm or $\kappa=21$~GHz as inferred from the quality factor. In addition to the fundamental mode, one can see the prominent second order transverse mode, and in between the odd first order mode, which couples weakly to the emitter. 

\section{Results}

\begin{figure}
	\centering
	\includegraphics{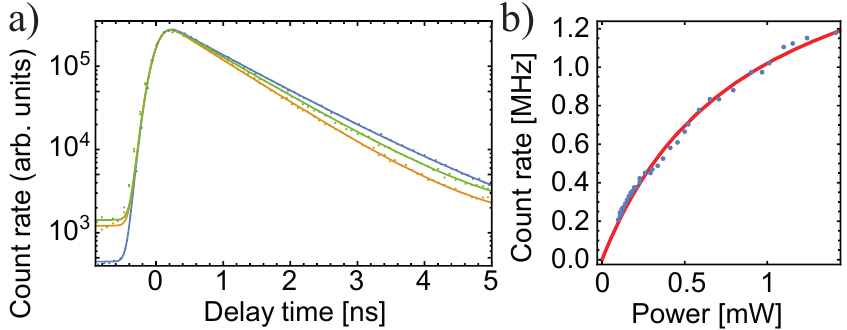}
	\caption{\label{fig:sat} (a) Lifetime measurement for ND3; data points and fit (solid line) with exponential function starting at $t=0$ convoluted with system response function. Blue: free space. Orange: cavity ($ q = 5 $). Green: cavity ($ q = 6 $) (b) Cavity saturation measurement of ND4 (Blue: data, red: fit).
	}
\end{figure}

The Purcell enhancement of the spontaneous emission leads to a reduction of the excited state lifetime, which we investigate by time-correlated single photon counting under pulsed excitation at 690~nm. Exemplary traces are shown in Fig.~\ref{fig:sat}a. We measure the instrument response function to be a Gaussian with $ \sigma = 0.157 $~ns and convolute it with an exponential decay as a fit function to reproduce the lifetime data. We perform such measurements both in free space and in the cavity at the longitudinal mode orders $q=5,6,7$. In the latter case, we stabilize the cavity on resonance by a piezo actuator and a software algorithm that maximizes the count rate. The polarization of the excitation light is in all cases chosen to match the projection of the dipole orientation in the plane of the mirror.
\begin{table}
	\caption{\label{tab:life}
		Measured lifetimes in free space ($ \tau_0 $) and in the cavity ($ \tau_c $: lowest reachable order, $ \tau_{c,2} $: following order). The error is obtained from different fitting methods (see appendix \ref{sec:lifetime}).}
	\begin{ruledtabular} 
		\begin{tabular}{cccccc}
			ND & $\tau_{0}$ {[}ns{]} & $\tau_{c}$ {[}ns{]} & $\tau_{c,2}$ {[}ns{]} & $\tau_{0}/\tau_{c}$\tabularnewline
			\hline 
			1 & $0.58\pm0.02$ & $0.46\pm0.02$ & $0.51\pm0.02$ & $1.28\pm0.01$\tabularnewline
			2 & $2.54\pm0.24$ & $1.75\pm0.04$ & $1.97\pm0.10$ & $1.45\pm0.14$\tabularnewline
			3 & $1.03\pm0.01$ & $0.85\pm0.01$ & $0.92\pm0.01$ & $1.21\pm0.01$\tabularnewline
		\end{tabular}
	\end{ruledtabular}
\end{table}
We observe a lifetime reduction in the cavity, and the effect is larger for smaller mirror separation, i.e. smaller mode volume. Table \ref{tab:life} summarizes the lifetime measurements taken for three different emitters and shows that lifetime reductions $ \Delta\tau / \tau_0 $ between 17\% and 31\% are observed.
The lifetime of the narrow emitters is in many cases found to be shorter than the reported value of 1 to 1.7~ns \cite{Rogers14,Arend16} in unstrained bulk diamond, which we also observe for SiV ensembles in this sample. We find a large spread of lifetimes \cite{Neu11}, which may originate from different strain in the nanodiamonds and a varying contribution of non-radiative decay \cite{Feng93}.
The ratio of lifetimes in free space, $ \tau_0 $, and the cavity, $ \tau_c $, depends on the Purcell factor and the quantum efficiency $\mathrm{QE}=\gamma_r/(\gamma_r+\gamma_{nr})$, where $ \gamma_r $ is the radiative and $ \gamma_r $ the non-radiative decay rate: $ \tau_0/\tau_c = C \, \mathrm{QE} + 1 $. For an unknown $ \mathrm{QE} < 1 $ as expected for SiV centers \cite{Riedrich14,Neu12}, we can thus not infer $C$ directly from the lifetime change. 


\begin{table}
	\caption{\label{tab:sat}
		Results of saturation measurements.  $I_{m,fs}^{\infty}$ ($I_{m,c}^{\infty}$): Saturation count rate in free space (cavity). $I_{fs}^{\infty}$ ($I_{c}^{\infty}$): Photon emission rate in free space (cavity) calculated from count rates. All rates in MHz. $ C_{th} $: maximal theoretical Purcell factor. $C_{\mathrm{exp}}=I_{c}^{\infty} / I_{fs}^{\infty}$. }
	\begin{ruledtabular}
		\begin{tabular}{ccccccc}
			ND & $I_{m,fs}^{\infty}$  & $I_{m,c}^{\infty}$  & $I_{fs}^{\infty}$ & $I_{c}^{\infty}$ & $C_{th}$ & $C_{\mathrm{exp}}$\tabularnewline
			\hline 
			1 & $0.092\pm0.007$ & $0.106\pm0.009$ & $0.57$ & $2.64$ & 9.1 & $4.6\pm0.5$\tabularnewline
			2 & $2.62\pm0.34$ & $1.21\pm0.33$ & $16.3$ & $28.6$ & 6.1 & $1.8\pm0.5$\tabularnewline
			4 & $1.07\pm0.40$ & $1.78\pm0.13$ & $6.6$ & $25.3$ & 9.9 & $3.8\pm1.4$\tabularnewline
			5 & $0.38\pm0.04$ & $1.65\pm0.90$ & $2.35$ & $21.6$ & 11.4 & $9.2\pm5.1$\tabularnewline
		\end{tabular}
	\end{ruledtabular}
\end{table}

To quantify the cavity enhancement of the emission and to obtain an estimate for the achieved Purcell factor, we compare the photon emission rate in free space and in the cavity under saturation conditions. In free space, the emission rate is given by $ I^\infty_{fs} = \gamma_r n_2^\infty $, where $ n_2^\infty  $ is the equilibrium population of the excited state. The emission rate into the cavity mode is $ I^\infty_c = C \gamma_r n_2^\infty$, and the ratio between the rates directly yields the Purcell factor, $C_{\mathrm{exp}}= I^\infty_c / I^\infty_{fs}$, independent of the quantum efficiency.

Experimentally, we measure the saturation count rate both in free space and in the cavity, and use the knowledge of collection and detection efficiencies to calculate back to the respective emission rates. An example for the saturation in the cavity is shown in Fig.~\ref{fig:sat}b, where we have again optimized the polarization of the excitation light. The measured rate $ I_m(P) $ as a function of the excitation power $ P $ can be described as $ I_m(P) = \frac{P I^\infty_m}{P+P_\mathrm{sat}} + a_{bg} P $ with $ I^\infty_m $ the count rate in the limit of large $ P $, $ P_\mathrm{sat} $ the saturation power, and $ a_{bg} P $ a linear term describing the contribution from background fluorescence. We find $a_{bg}= 62~(40)\times 10^3$~cts/(s\,mW) in free space (in the cavity) for ND4. The obtained values for $ I^\infty_m $ are given in table \ref{tab:sat}, where the errors are from the uncertainty of the fit. We observe saturation count rates at the detector of up to $I_{m,c}^\infty=1.78\times 10^6$cts/s, corresponding to a peak spectral density of $2I_{m,c}^\infty/(\pi \kappa)= 54\times 10^3$cts/(s\,GHz). The peak spectral density is more than 20-fold larger than in earlier room-temperature experiments \cite{Neu11,Englund10,Albrecht13,Riedrich14,Kaupp16}.

To obtain the photon emission rates (emission rate into a solid angle of $ 4\pi $ in free space, $I^\infty_\mathrm{fs}$, and into the cavity mode, $ I^\infty_c$) from measured count rates, we account for the collection and detection efficiency in both cases. In free space, the light is collected with an NA 0.55 objective, and the emission is enhanced and directed due to the presence of the mirror. In the cavity case, we consider the outcoupling efficiency through the fiber mirror $\eta_c$ and the mode matching between the cavity and fiber modes $\epsilon=0.47$. Values for $\eta_c$ vary between 20\% and 38\% for the investigated emitters, limited by scattering and absorption of the nanodiamonds (see appendix \ref{sec:countrate} for more details). With an improved sample, this loss channel can be avoided. The obtained photon emission rates are given in table \ref{tab:sat}.
In free space, we infer $I^\infty_\mathrm{fs}$ of up to 16~MHz for ND2, while in the cavity, we find a rate $ I^\infty_c$ of more than 28~MHz. From the ratio of the two rates, we obtain values for the Purcell factor of up to $ C_\mathrm{exp} =9.2$ (ND5). This corresponds to an efficiency to collect the emitted photons with the cavity mode of $ \beta=90\% $. The stated errors stem from the uncertainties of the fit of the saturation curves, and do not include further uncertainties. We can also infer the enhancement of the spectral density compared to free space emission, $C_\mathrm{exp} Q_c/Q_\mathrm{em}$, yielding a value of 237 for ND5.

We compare $C_\mathrm{exp}$ to the expected maximal Purcell factor $C_\mathrm{th}$ as calculated from $Q_c,Q_\mathrm{em}$, and $V_m$ for the respective emitters. In the calculation, we obtain $Q_\mathrm{em}$ from the linewidth of the measured emission spectra and calculate $Q_c$ for the respective emission wavelength. Furthermore, we assume optimal coupling conditions, i.e. $\eta_E\equiv\left|\frac{\vec{\mu}\vec{E}}{\mu E_0}\right|^2=1$ as an upper bound. The ideal value is almost reached for ND5, but the experimental results stay well below the ideal enhancement for the other three emitters (see table II). This is explained by an unfavorable position of the emitter within the crystal, or a non-ideal dipole orientation, leading to $\eta_E<1$. As those factors also enter the collection efficiencies, the values of $C_\mathrm{exp}$ contain additional uncertainties.

From $C_\mathrm{exp}$ and the lifetime change $ \tau_0/\tau_c $, we can coarsely estimate the quantum efficiency for ND1 and 2, and find $\mathrm{QE} \approx 7\%$ and 25\%, respectively. The former is comparable to previously published values \cite{Riedrich14,Neu12}, the latter appears inconsistently high (see appendix \ref{sec:rates}). A low QE can also explain the low emission rate of ND1 despite its short lifetime.
Notably, the Purcell effect leads to an increased QE. One finds that the QE in the cavity, $ \mathrm{QE}_c $, relates to the free-space QE via $ \mathrm{QE}_c=(C+1)/(C+1/\mathrm{QE}) $, such that e.g. for ND1, the QE increases from 7\% to 30\% in the cavity.
The overall device efficiency, which states the probability to obtain a photon in the fiber after excitation of the emitter, is given by $\beta_\mathrm{tot}= \mathrm{QE}_c \beta \eta_c \epsilon$, and we obtain $\beta_\mathrm{tot}= \mathrm{QE}_c\times 16\%$ for ND5.


\begin{figure}
	\centering
	\includegraphics{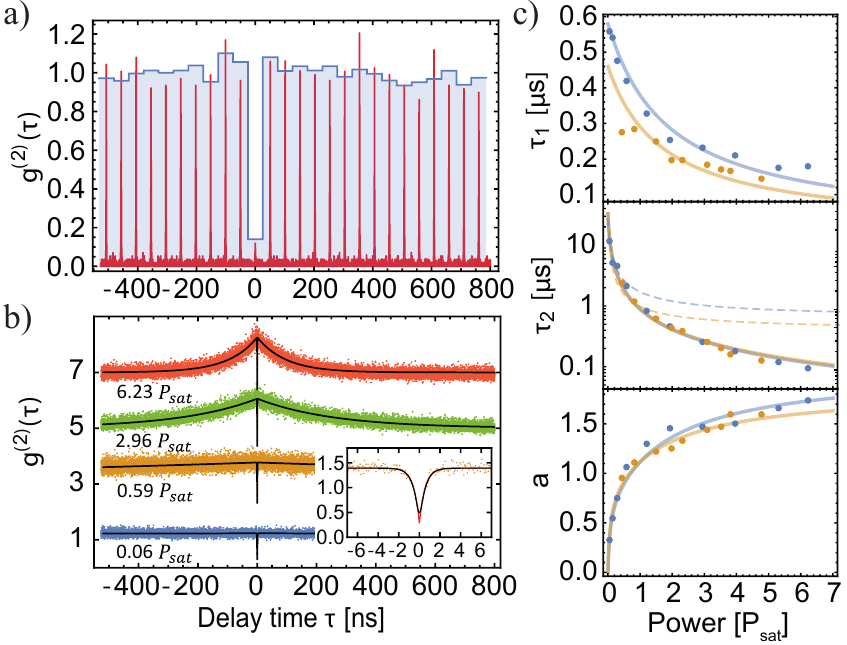}
	\caption{\label{fig:g2} (a) Free space $ g^{(2)} $ measurement of ND1 using pulsed excitation (50~ps pulse duration, 20 MHz repetition rate, wavelength 690 nm). Red curve: data.  Blue histogram: Integrated over each peak, background subtracted. (b) Free space $ g^{(2)} $ measurement (cw excitation at 690 nm) of ND1 for selected excitation powers. Inset: Zoom into data for $ 0.17 P_{sat} $.  Black solid lines: Fit with $ g^{(2)} $ function convoluted with system response function. Orange solid line: $ g^{(2)} $ function without convolution. (c) Fit parameters as a function of power and fit with power-dependent deshelving model (solid lines). Dashed: Fit with model from \cite{Neu12}.
	}
\end{figure}

To prove that the emitters show single photon emission, we measure the intensity correlation function $ g^{(2)}(\tau) $ with a Hanbury-Brown-Twiss (HBT) setup. Figure \ref{fig:g2}a shows a pulsed $ g^{(2)} $ measurement yielding a value $ g^{(2)}(0) = 0.14 $ for zero time delay (see appendix \ref{sec:data} for data of the other emitters).
For continuous wave excitation (Fig.~\ref{fig:g2}b), we observe a power-dependent photon bunching for intermediate time delays consistent with previous observations (see e.g. \cite{Neu12}), which can be attributed to a meta-stable shelving state \cite{Gali13}. The dynamics of such a three level system is described by
\begin{equation}\label{eq:g2}
g^{(2)}(\tau) = 1-(1+a) e^{-|\tau|/\tau_1} + a e^{-|\tau|/\tau_2},
\end{equation}
which fits the data well.
Since the shelving state might significantly reduce the achievable excited state population $n_2^\infty$ and emission rate, it is important to understand the internal rate dynamics. Therefore, we measure the $ g^{(2)} $ function for various powers both in the cavity and in free space and fit with Eq.~\ref{eq:g2} including uncorrelated background. The fit parameters $ \tau_1 $, $ \tau_2 $, and $ a $ are given in Fig.~\ref{fig:g2}c as functions of excitation power. We find that the antibunching time constant $ \tau_1 $ is smaller in the cavity, as expected. Note that $ \tau_1(0) $ is equivalent to the spontaneous emission lifetime of the system being measured.
We observe a strong power dependence of the bunching time constant $ \tau_2 $ \cite{Neu12}, which is equally large in the cavity and in free space. However, the proposed model for an intensity-dependent deshelving (\cite{Neu12}, dashed line) does not accurately describe the data. Rather than approaching a finite value, $ \tau_2 $ converges to zero for large powers. This can be explained by a revised model, which allows linearly power dependent excitation to a higher lying state both from the exited state and the shelving state (see appendix \ref{sec:rates}). An ionization process could explain these dynamics. For the emitter studied, the model yields a rather large value $n_2^\infty=0.34$, and the short $ \tau_2 $ at high power indicates the possibility of high repetition rates for pulsed excitation.

\section{Conclusion}

We have shown significant Purcell enhancement of the single photon emission of SiV centers, achieving high efficiency ($\beta=90\%$), a high photon rate coupled into a single-mode fiber (4.1 MHz), and a narrow linewidth (21 GHz) at room temperature. Several emitters show excited state lifetimes below 1~ns and bunching time constants that decay quickly with power, such that pulsed excitation at GHz rates is possible. We have introduced a revised rate model to accurately describe the power dependent dynamics of the SiV center.
The reported experiments were limited by properties of the sample, such as excessive scattering and absorption loss, photobleaching of the emitters after excitation times ranging from minutes to weeks, and a moderate quantum efficiency. On the cavity side, smaller mode volume and higher quality factor cavities have been fabricated and promise Purcell factors of about 40 for 1~nm emitter linewidth, and outcoupling efficiencies up to $\eta_c=97\%$ are achievable for small nanodiamonds.
For an improved SiV sample with $\mathrm{QE}=30\%$ \cite{Riedrich14} coupled to an optimized cavity, our approach has the prospect to achieve a device efficiency of $\beta_\mathrm{tot}=90\%$ and yield single photon rates beyond 1 GHz. Furthermore, using a high-$Q$ cavity, spectral purity can be improved to a level where indistinguishable single photons could be produced under ambient conditions \cite{Grange15}, meeting the challenging requirements for all-optical quantum computation \cite{Knill00,Kok07}.

\appendix

\section{Cavity properties}\label{sec:cavity}
\begin{figure}
	\centering
	\includegraphics{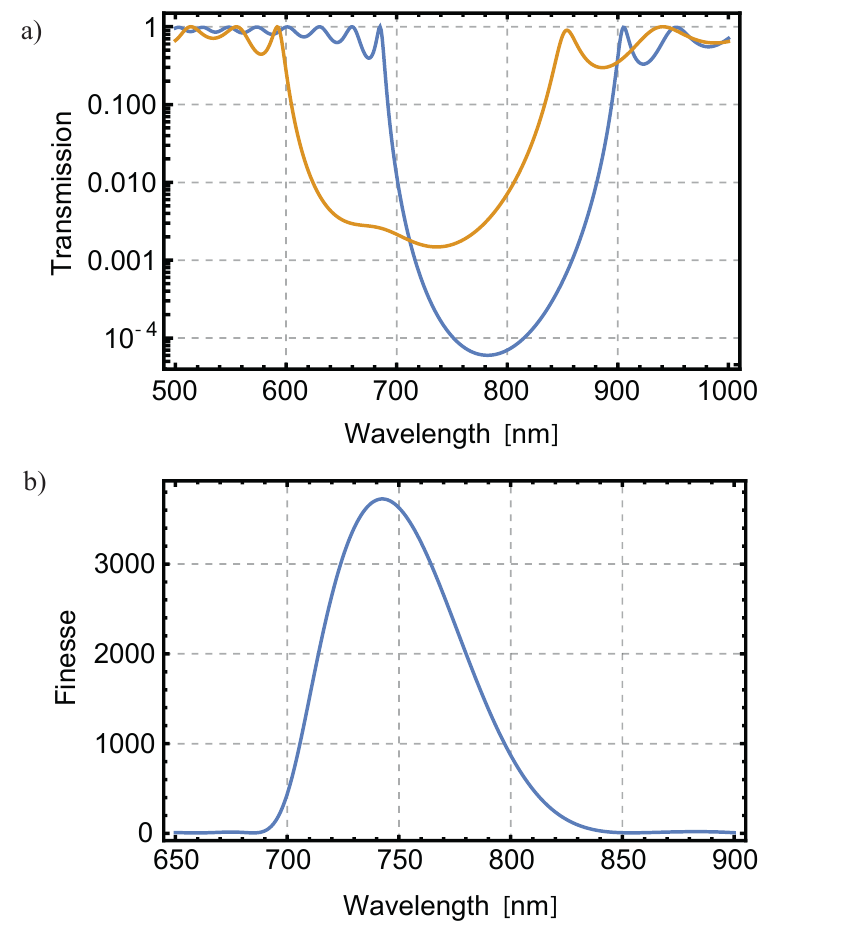}
	\caption{\label{fig:trans}(a) Computed mirror transmission of the fiber (yellow) and the planar mirror (blue). (b) Computed finesse assuming the above mirror transmissions and 40~ppm absorption losses. 
	}
\end{figure}

The fiber is coated with a dielectric mirror with a transmission of 1500~ppm at a center wavelength of 740~nm and 2500~ppm at the excitation wavelength of 690~nm, such that no excitation light enters the fiber to avoid fiber fluorescence. The planar mirror has a coating centered at 780~nm with a transmission of 60~ppm (200~ppm at 740~nm) and is designed to yield an electric field maximum 30~nm above the mirror surface. The coating is almost transparent at 690~nm, and we focus the excitation light into the cavity with an aspheric lens through the planar mirror. The asymmetry in transmission leads to about 90\% of the fluorescence light being emitted into the fiber, which is our collection channel. Figure \ref{fig:trans}a shows a simulation of the fiber and mirror coating  using the transfer matrix method.
Assuming a total absorption loss at the mirrors of 40~ppm, this yields the cavity finesse as shown in Fig. \ref{fig:trans}b. The measured finesse of 2000 at 780~nm coincides well with the computation such that it is a fair assumption that the computed value of 3750 at 740~nm is also reached.

The mirror is held by a gimbal mount for angular alignment and can be reproducibly moved in all three directions with a nanopositioning stage.

\section{Sample scans}\label{sec:scans}

\begin{figure}
	\centering
	\includegraphics{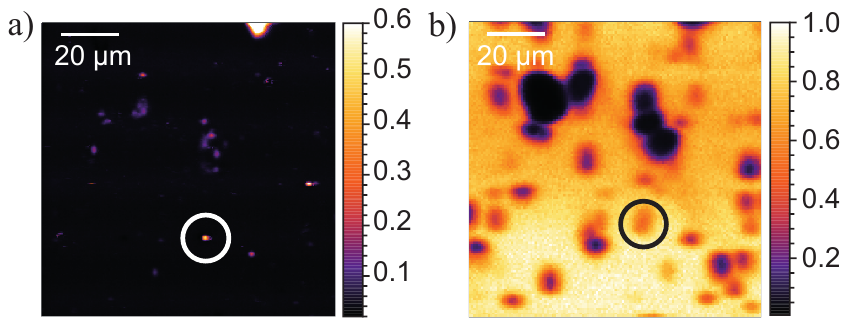}
	\caption{\label{fig:scans}(a) Confocal fluorescence map showing the count rate in MHz. The marked spot is a narrow line emitter. (b) Cavity transmission scan of the same area (transmission normalized to 1) for a cavity length of about $10~\mu$m. The marked spot is the same location as in a). 
	}
\end{figure}

The mirror can be shifted between an objective and the fiber using a single nanopositioning stage (attocube ECS3030), such that the same area can be investigated both with a confocal microscope and in the cavity. Figure \ref{fig:scans}a shows a fluorescence map containing a narrow line single emitter (circle). Then, a scanning cavity microscopy map of the same area is recorded (see Fig. \ref{fig:scans}b), where nanodiamonds and other nanoparticles show up as dark spots. A calibration of the offset between the confocal focus and the fiber enables us to quickly switch between the two observation methods and easily find a pre-characterized emitter in the cavity. The extinction induced by a chosen emitter enables optimization of the spatial overlap with the cavity mode.

\section{Extended data}\label{sec:data}

In this paragraph, we show the complete set of emitter properties as extracted from the free space spectra and an exemplary comparison of saturation measurements in free space and in the cavity.
\begin{table}
	\caption{\label{tab:emitters}
		Overview of considered nanodiamonds. $ \lambda $: center wavelength of line. $ \delta $: FWHM of emission spectrum. $ Q_\mathrm{em} = \lambda/\delta $. The finesse at $ \lambda $ was used to compute $ \Qeff $ and $ C_\mathrm{eff} $. $ g^{(2)}(0) $: Minimum of $ g^{(2)} $ function obtained from pulsed measurement. Exception: Value for ND5 was obtained from cw measurement.}
	\begin{ruledtabular} 
		\begin{tabular}{cccccc}
			ND & $\lambda$ {[}nm{]} & $\delta$ {[}nm{]} & $Q_{\mathrm{em}}$ & $C_{\mathrm{eff}}$ & $ g^{(2)}(0) $\tabularnewline
			\hline
			1 & 758.9 & 1.4 & 542 & 9.1 & 0.14\tabularnewline
			2 & 754.3 & 2.1 & 359 & 6.1 & 0.22\tabularnewline
			3 & 758.0 & 3.0 & 253 & 4.4 & 0.44\tabularnewline
			4 & 758.6 & 1.3 & 584 & 9.9 & 0.22\tabularnewline
			5 & 752.4 & 1.1 & 684 & 11.4 & 0.54\tabularnewline
			6 & 736.9 & 1.3 & 567 & 9.1 & 0.19\tabularnewline
		\end{tabular}
	\end{ruledtabular}
\end{table}
\begin{figure}
	\centering
	\includegraphics{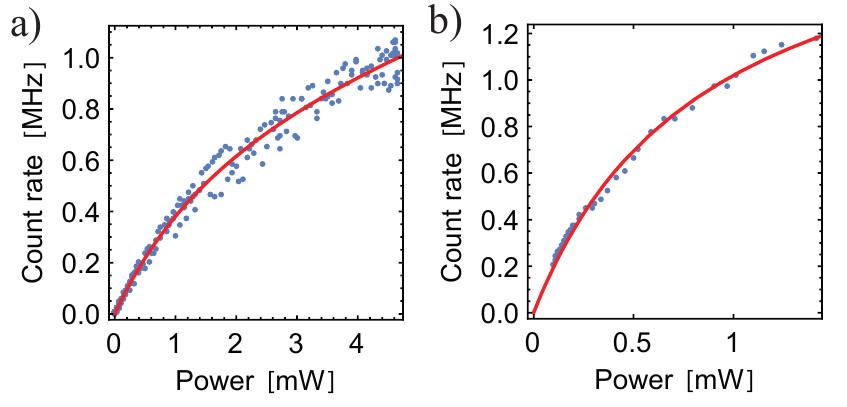}
	\caption{\label{fig:sat2}(a) Free space saturation measurement of ND4. Blue: data. Red: fit. (b) Cavity measurement of ND4. 
	}
\end{figure}

Table \ref{tab:emitters} gives an overview of all emitters considered including their center wavelength and linewidth as obtained from a Lorentzian fit.  The cavity finesse was calculated for the wavelengths of all emission lines from the coating parameters. Together with the width $ \delta $ of the lines, $ Q_\mathrm{em} = \lambda/\delta $ and a theoretical value for $ C_\mathrm{eff} $ were computed and are also given in table \ref{tab:emitters}. It includes the branching factor  $\zeta=0.8$ of the ZPL. The values are upper bounds for the case of optimal position and dipole orientation of the emitter, i.e. $\eta_E\equiv\left|\frac{\vec{\mu}\vec{E}}{\mu E_0}\right|^2=1$. The minimum values of the $ g^{(2)} $ function are obtained using pulsed excitation at 690~nm (pulse length 50 ps) strongly saturating the emitters. They are thus to be seen as an upper bound as the background increases linearly with power and therefore contributes more when $ P > P_{sat} $ increases. ND5 is an exception as it underwent photobleaching before pulsed measurements could be taken. Here, we give the minimum of the fit to a continuous wave 690~nm $ g^{(2)} $ measurement. Although being slightly larger than 0.5, the data is consistent with a single quantum emitter when taking into account the background fluorescence during this measurement (about 30\%). 

Figure \ref{fig:sat2} shows a comparison of a saturation measurement in free space and the cavity together with a fit containing a linear background. We find that the background is dominated by broadband fluorescence from the crystal, which greatly varies between crystals, and in the cavity, the background is suppressed. The given count rate is the actually detected rate.

\section{Fitting of lifetime data}\label{sec:lifetime}
For obtaining the excited state lifetime, we apply different fitting methods to the data and give the half min-max deviation as a conservative error of the fit (see table \ref{tab:life}). First, we fit the whole trace, i.e. an exponential decay function starting at delay time zero and including a constant background, convoluted with the system response function. For some emitters, a second exponential function has to be included to account for a fast decaying background. This is mostly necessary for the free space time trace. Next, only the decay is fitted to avoid systematic errors due to the positioning of the zero time delay. Last, we fit the section of exponential decay only.

\section{Calculation of the photon emission rate from the count rate}\label{sec:countrate}

To obtain the Purcell factor, one has to compare the photon emission rate into the cavity mode to the free space photon emission rate into $ 4\pi $. In order to trace back to those rates from the count rates on the detector, the collection and detection efficiency have to be accounted for.

The detection efficiency of the APDs is $ \eta_{det} = 70\% $ for the emitter wavelengths. From a measurement, we find that $ \eta_{trans} = 64\% $ of the fluorescence light is transmitted through the optics (including filters) from the first lens to the detector. It is a fair assumption that this value is the same for both the free space and cavity situation as the light travels the same path.

As the nanodiamonds reside on the planar mirror even in the confocal measurement, the local density of states is different as compared to free space. Here, we are not interested in the collected fraction of actually emitted light, but rather in the fraction of light collected as compared to free space emission into $ 4\pi $. To obtain this value, we compute the complex reflectivity $ r_{s,p} $ of the dielectric mirror stack for both s- and p-polarization using the transfer matrix method \cite{Furman92,Hood01}. The power radiated by a dipole is given by
\begin{equation}
P_s(\alpha,\theta,\phi) = \frac{3}{8\pi} \sin^2 \theta \sin^2 \phi
\end{equation}
for s-polarization and
\begin{equation}
P_p(\alpha,\theta,\phi) = \frac{3}{8\pi} (\cos \theta \sin \alpha + \sin \theta \cos \alpha \cos \phi)^2
\end{equation}
for p-polarization, where $ \alpha $ the azimuthal angle between the optical axis and the emission direction, $ \phi $ the polar angle, and $ \theta $ the angle between the dipole and the optical axis. The total power emitted by the dipole on the mirror into a certain direction normalized to free space power is the interference of the directly emitted and reflected light in both polarizations \cite{Lukosz79}:
\begin{multline}
P_{em}(\alpha,\theta,\phi) = \sum_{i\in \{s,p\}} \left| 1+|r_i|\exp(2k\cos\alpha z_0 + \arg r_i) \right| ^2 \\ \cdot P_i(\alpha,\theta,\phi)
\end{multline}
The term $ 2k\cos\alpha z_0 $ with $ k $ the wave number, is an additional phase acquired if the dipole has a distance $ z_0 $ from the surface of the mirror. Integrating this quantity over the solid angle given by the numerical aperture ($ \mathrm{NA} = 0.55 $) yields the desired collection efficiency:
\begin{equation}
\eta_{coll}(\theta) = \int_{0}^{\arcsin \mathrm{NA}} \int_{0}^{2\pi} P_{em}(\alpha,\theta,\phi) \sin\alpha \mathop{}\!\mathrm{d}\phi \mathop{}\!\mathrm{d}\alpha
\end{equation}
For a dipole oriented parallel to the mirror surface, we obtain $ \eta_{coll}(\pi/2) = 45\% $, for perpendicular orientation $ \eta_{coll}(0) = 8\% $.
As we choose very bright emitters for the experiment, there is a bias towards those with optimal, i.e. parallel, dipole orientation. We therefore use $ \eta_{coll}(\pi/2) $ as a good estimate for the collection efficiency. This value carries an additional uncertainty, as we do not know the exact position of the dipole within the crystal. Again, we assume that we preselect diamonds in which the dipole resides close to the field maximum as these will appear brighter. The objective has a transmission of about $ \eta_{obj} = 80\% $ for the fluorescence wavelength.

In summary, the free space photon emission rate into $ 4\pi $ can be calculated by
\begin{equation}
I_{fs}^\infty = I_{m,fs}^\infty /(\eta_{det}  \eta_{trans}  \eta_{coll}  \eta_{obj}).
\end{equation}

\begin{table}
	\caption{\label{tab:eff}
		$ T/T_0 $: extinction. $ \eta_c $: outcoupling. $C_{\mathrm{exp}}$: experimental Purcell factor. $ \beta = C/(C+1)$: fraction of total emission into the cavity mode. $\beta \eta_c$: fraction of emission coupled out of the cavity.}
	\begin{ruledtabular} 
		\begin{tabular}{cccccc}
			ND & $T/T_{0}$ & $\eta_c$ & $C_{\mathrm{exp}}$ & $\beta$ & $\beta \eta_c$\\
			\hline 
			1 & 0.25 & 0.21 & $4.2\pm0.5$ & 0.81 & 0.17\\
			2 & 0.23 & 0.21 & $1.8\pm0.5$ & 0.64 & 0.13\\
			5 & 0.45 & 0.35 & $3.8\pm1.4$ & 0.79 & 0.28\\
			6 & 0.50 & 0.38 & $9.2\pm5.1$ & 0.90 & 0.34\\
		\end{tabular}
	\end{ruledtabular}
\end{table}

In the cavity case, we are interested in what part $ \eta_c $ of the emitted light is coupled out through the fiber mirror, which is our collection channel. It is given by the transmission of this mirror divided by all losses:
\begin{equation}
\eta_c = T_f/(T_f+T_p+A+L),
\end{equation}
where $ T_f $ ($ T_p $) is the transmission of the fiber (planar) mirror, $ A $ the absorption and scattering losses of both mirrors, and $ L $ the extinction losses due to absorption and scattering of the nanocrystal. $ L $ can be calculated from the extinction $ T/T_0 $, which is given in table \ref{tab:eff}. To obtain the extinction, we measure the cavity transmission and normalize it to the transmission of the empty cavity. The cavity transmission is given by
\begin{equation}
T=  \frac{4T_s T_f}{(T_s+T_f+A+L)^2},
\end{equation}
so $ L $ can be determined by solving the following equation:
\begin{equation}
(T_0/T) (T_s+T_f+A)^2 = (T_s+T_f+A+L)^2
\end{equation}

As the light is collected through the fiber, one has to take into account the mode matching between the cavity mode and the mode guided by the fiber, whose mode field radius $ w_f $ is 2.5\um. For the situation that the modes are coaxial (for a well-centered fiber profile and good angular alignment), the power coupling efficiency can be computed as 
\begin{equation}
\epsilon = \frac{4}{\left(\frac{w_f}{w_0}+\frac{w_0}{w_f}\right)^2+\left(\frac{s \lambda}{\pi w_0 w_f}\right)^2},
\end{equation}
where $ w_0 $ is the mode waist and $ s = d + d_{mirror} $ is the optical distance between the two mode waists \cite{Joyce84,Hunger10b}. The latter is composed of the geometric mirror separation $ d $ and the optical thickness of the fiber's dielectric mirror stack $ d_{mirror} $. The mode waists are calculated for all cavity lengths by optimizing the Gaussian mode for the given fiber profile (for details see \cite{Kleckner10,Benedikter15}). For the lowest achievable cavity length $ d_\mathrm{eff} = 5 \lambda/2 $, we obtain $ w_0 = 1.00 $\um.  This leads to a mode matching $ \epsilon = 46.6 \% $. Note that this value is an upper boundary as already a slight misalignment of the fiber profile with respect to the fiber core can significantly reduce the mode matching. Therefore, using $ \epsilon $ for calculating the photon emission rate yields a conservative estimate. At the glass-air-interface at the outcoupling port of the fiber, another 4\% get lost, and we get an additional factor $ \eta_{fiber} = 96\% $.

The photon emission rate into the cavity can then be calculated from the count rate as
\begin{equation}
I_{cav}^\infty = I_{m,cav}^\infty /(\eta_{det} \eta_{trans}  \eta_c  \eta_{fiber} \epsilon).
\end{equation}

Due to the large uncertainties of some of the factors, the obtained photon emission rates and Purcell factors are an estimate.

Table \ref{tab:eff} also gives the efficiency $ \beta=C/(C+1) $, which is the fraction of the total emission into the cavity mode. We obtain values up to 90\%. The fraction of the light actually coupled out of the cavity $ \beta \eta_c $ is given in the last column. Note that this could be significantly improved by choosing a sample with less extinction, i.e. smaller crystal size and better crystal quality.

The photon emission rates given only include the emission into the cavity mode, i.e. $ I^\infty_{cav}=\beta I^\infty_\mathrm{tot} $ where $ I^\infty_\mathrm{tot} $ would be the total rate into $ 4\pi $. As $ I^\infty_\mathrm{tot} = (C+1) I^\infty_{fs} $, the ratio of cavity and free space emission rate yields just $ C $:
\begin{equation}
I^\infty_{cav}=\beta I^\infty_\mathrm{tot}=\beta (C+1) I^\infty_{fs}=C I^\infty_{fs}
\end{equation}

\section{Theoretical Purcell factor calculation}\label{sec:purcell}

The theoretical Purcell factors as stated in Table \ref{tab:emitters} are calculated as 
\begin{equation}\label{eq:Ceff}
C_\mathrm{eff} = \frac{3(\lambda/n)^3}{4\pi^2}\frac{\Qeff}{V_m} \zeta \left|\frac{\vec{\mu}\vec{E}}{\mu E_0}\right|^2
\end{equation}
with
\begin{equation}
\Qeff =(Q_c^{-1}+Q_\mathrm{em}^{-1})^{-1}
\end{equation}
being the effective quality factor, $ \zeta = 80\%$, and assuming ideal dipole location and orientation, i.e. $\left|\frac{\vec{\mu}\vec{E}}{\mu E_0}\right|^2=1$. The mode volume $ V_m $ is calculated using the effective cavity length and the optimized mode waist. $ Q_c $ is adjusted individually for all emitters as the additional extinction loss $ L $ alters the finesse on the emitter as compared to the bare cavity. This is however not very critical, as $ Q_c \gg Q_\mathrm{em} $ still holds. The most important quantity is $ Q_\mathrm{em} = \lambda/\delta\lambda $ and is obtained from the measured linewidth $ \delta\lambda $. The resulting effective Purcell factors set an upper limit for the achievable experimental values for the case the dipole is oriented parallel to the mirror surface and resides in the maximum of the electric field. In reality, this does not have to be the case and the effects on the Purcell factor are discussed here.

\begin{figure}
	\centering
	\includegraphics{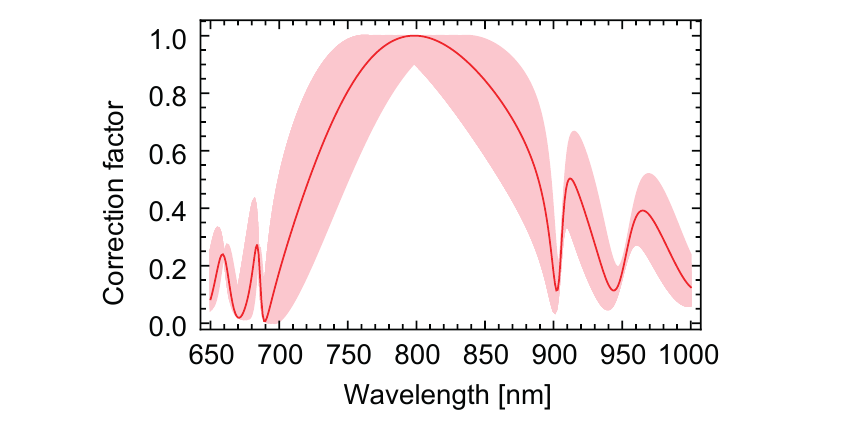}
	\caption{\label{fig:factor}Correction factor for Purcell factor due to the emitter's position relative to the electric field maximum. Red line: for a color center residing in the middle of a 100~nm large diamond. Shaded area: uncertainty of the correction factor due to the unknown position of the emitter. 
	}
\end{figure}

The planar mirror, on which the sample is placed, is a dielectric Bragg reflector at a central wavelength of 780~nm. It is capped with a spacer layer of $ \mathrm{SiO_2} $ such that the field maximum for 780~nm lies 30~nm above the surface. 
Using the complex reflectivity $ r $ of the planar mirror, $ |1+r^2| $ gives the field intensity normalized to the value without a mirror. Dividing this by the intensity maximum, we obtain a correction factor $|E/E_0|^2$ for the Purcell factor, which is plotted for different wavelengths in Fig. \ref{fig:factor}. The red line is computed assuming the color center resides in the middle of a 100~nm large diamond. The shaded area gives the uncertainty of the correction factor due to the unknown position of the emitter. As an example, at a wavelength of 754~nm, the correction factor can range from 56\% to almost 100\%. So an unfortunate position of the color center within the crystal can decrease the Purcell factor to half its ideal value.

The second unknown quantity is the orientation of the dipole moment. By maximizing the count rate, we align the polarization of the electric field with the in-plane component of the dipole moment. Therefore, only the out of plane component remains unknown, which can be characterized by an angle $ \phi = \angle(\vec{\mu},\vec{E}) $. Considering equation \ref{eq:Ceff}, we find that $ C \propto \sin^2 \phi $. Therefore, the Purcell factor goes down to zero for $ \vec{\mu} \perp \vec{E} $. But as the excitation goes likewise down, emitters with unfavorable orientation of $ \vec{\mu} $ are not bright and are likely not to be preselected.

\section{Rate dynamics}\label{sec:rates}

We assume a level scheme as depicted in Fig. \ref{fig:levels}a, where 1 is the ground, 2 the excited, and 3 the shelving state. The rate $ k_{12} $ is linearly dependent on the excitation power as we excite off-resonantly with subsequent fast relaxation into the phononic ground state. A three level system with otherwise constant rates leads to a constant bunching time constant $ \tau_2 $ \cite{Neu12}, which is in contradiction with observation. We attribute this deviation to excitation from levels 2 and 3 to some higher lying state or the valance band. The excitation rates are also assumed to be linearly dependent on power. All deexcitation rates are constant. The cavity is resonant with transition $ 2\to 1 $. For easier mathematical treatment, an equivalent level scheme is presented in Fig. \ref{fig:levels}b leading to the same dynamics. Here, green arrows indicate constant rates, red arrows rates that are linearly dependent on power and the blue rate $ k_{32} $ has both a linear and a constant term:

\begin{figure}
	\centering
	\includegraphics{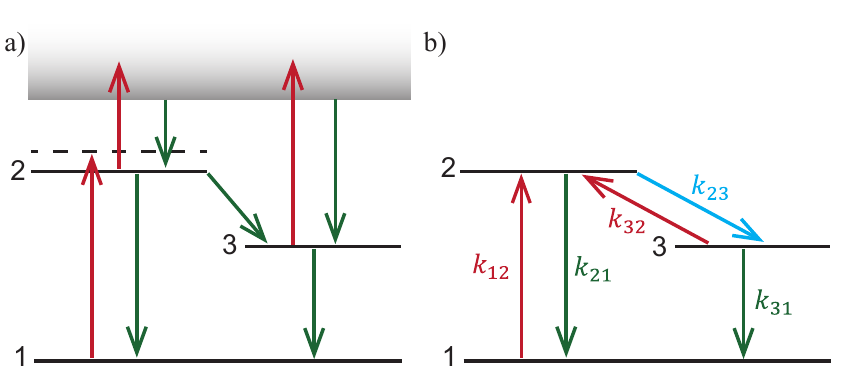}
	\caption{\label{fig:levels} (a) Proposed level scheme. Green: constant rates. Red: linearly power dependent rates. (b) Equivalent three level system. Blue: rate with linear and constant term. 
	}
\end{figure}

\begin{align}\label{eq:rates}
k_{12} &= \sigma P\\
k_{32} &= d P\\
k_{23} &= e P + k_{23}^0,
\end{align}
where $ \sigma $, $ d $, and $ e $ are proportionality constants. The $ g^{(2)} $ function for this three level system is given by
\begin{equation}\label{eq:g2b}
g^{(2)}(\tau) = 1-(1+a) e^{-|\tau|/\tau_1} + a e^{-|\tau|/\tau_2},
\end{equation}
with
\begin{align}
\tau_{1,2} &= 2/(A \pm \sqrt{A^2-4B}) \\
a &= \frac{1-\tau_2(k_{31}+k_{32})}{(k_{31}+k_{32})(\tau_2-\tau_1)}
\end{align}
and
\begin{align}
A &=  k_{12} + k_{21} + k_{23} + k_{31} + k_{32} \\
\begin{split}
B &= k_{23} k_{31} + k_{21} (k_{31} + k_{32})\\
&+ k_{12} (k_{23} + k_{31} + k_{32}).
\end{split}
\end{align}
Like in \cite{Neu12}, we express the constant rates by values for the parameters at zero power:
\begin{align}
k_{21} &= 1/\tau_1^0 - k_{23}^0 \equiv \gamma_r + \gamma_{nr}\\
k_{31} &= 1/\tau_2^0.
\end{align}
In the limit of large powers, both time constants converge to zero. The bunching parameter $ a $ is zero for zero power and reaches a finite value $ a^\infty $ for large powers. We use it for fitting instead of the proportionality constant $ e $ as the starting value is easier to choose:
\begin{equation}
e = \frac{-a^\infty d^2+ a^\infty d \sigma}{a^\infty d + \sigma}.
\end{equation} 
In Section III, we compare the model with data taken from ND1. We perform a global fit of six data sets for $g^{(2)}(\tau)$ taken at different excitation power for $ a $, $ \tau_1 $, and $ \tau_2 $ both in free space and in the cavity. The fit has eight free parameters all together ($ d $, $ d_c $, $ \sigma $, $ \sigma_c $, $ a^\infty $, $ a^\infty_c $, $ \tau_2^0 $, and $ k_{23}^0 $) as $ \tau_2^0 $ and $ k_{23}^0 $ are the same in both cases. From these parameters, one can calculate the equilibrium population of the excited state $ n_2^\infty $ and finally the total deexcitation rate $ \Gamma $:
\begin{align}
&n_2 = \frac{k_{12}(k_{31}+k_{32})}{k_{23}k_{31}+k_{21}(k_{31}+k_{32})+k_{12}(k_{23}+k_{31}+k_{32})}\nonumber\\
&n_2^\infty =\lim_{P\to\infty}n_2 = \frac{1}{1+e/d}\\
&\Gamma = n_2^\infty k_{21}
\end{align}
We obtain $ n_2^\infty = 34\% $ for ND1. In free space (in the cavity), we get $ k_{21} = 1.7 $~GHz (2.2~GHz) and $ \Gamma = 578 $~MHz (750~MHz), comparable to the values found in \cite{Neu12}. $ \Gamma $ includes non-radiative deexcitation, such that it should be possible to extract the QE by a comparison with the measured photon emission rate $ I_{fs}^\infty $. This yields a value for the QE of 0.1\% to 0.5\% depending on the orientation of dipole, significantly less than the 7\% estimated from the experimental Purcell factor and the lifetime change. The origin of this discrepancy remains unclear at this stage. The emitter might show different levels of blinking at intermediate timescales, or have jumped into another state with a different emission rate between the measurements (both are phenomena we occasionally observe).

\section{Sample properties}\label{sample}

\begin{figure}
	\centering
	\includegraphics{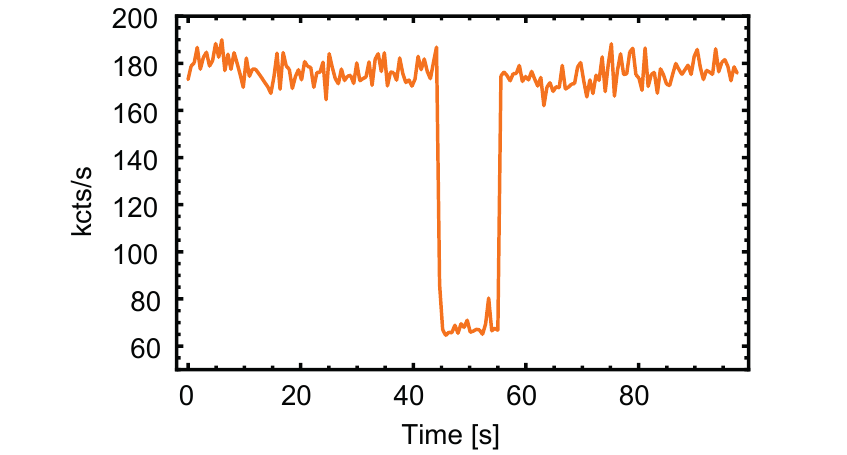}
	\caption{\label{fig:blinking} Fluorescence timetrace (count rate on detector) of single emitter featuring blinking.
	}
\end{figure}

As previously observed \cite{Neu12}, the used sample contains a small fraction of photostable emitters as well as emitters which feature blinking (see e.g. Fig. \ref{fig:blinking}) on different timescales (ranging from less than seconds to hours) and permanent photobleaching. The latter can occur after illumination times from seconds to several weeks and is more probable for higher excitation powers. This has prevented us from obtaining complete data sets for all emitters. New samples \cite{Li16}, however, promise better emitter photostability.

The large scattering and absorption losses reduces the number of useful emitters for investigation. It is therefore crucial to use smaller diamonds. Absorption can be caused by sp\textsuperscript{2} hybridized carbon at surfaces, grain boundaries and lattice defects. A higher crystal quality would thus reduce absorption and background fluorescence. This could for example be obtained by annealing of the nanodiamonds under oxygen atmosphere.

\begin{acknowledgments}
	We thank Roland Albrecht, Elke Neu, and Matthias Mader for contributions to the experiment. Fruitful discussions with John Rarity are acknowledged.
	The work has been funded by the European Union 7th framework Program under grant agreement no. 61807 (WASPS), the DFG Cluster of Excellence NIM, and the DFG project FOR1493. T. W. H\"ansch acknowledges funding from the Max-Planck
	Foundation.
\end{acknowledgments}


\begin{thebibliography}{51}%
	\makeatletter
	\providecommand \@ifxundefined [1]{%
		\@ifx{#1\undefined}
	}%
	\providecommand \@ifnum [1]{%
		\ifnum #1\expandafter \@firstoftwo
		\else \expandafter \@secondoftwo
		\fi
	}%
	\providecommand \@ifx [1]{%
		\ifx #1\expandafter \@firstoftwo
		\else \expandafter \@secondoftwo
		\fi
	}%
	\providecommand \natexlab [1]{#1}%
	\providecommand \enquote  [1]{``#1''}%
	\providecommand \bibnamefont  [1]{#1}%
	\providecommand \bibfnamefont [1]{#1}%
	\providecommand \citenamefont [1]{#1}%
	\providecommand \href@noop [0]{\@secondoftwo}%
	\providecommand \href [0]{\begingroup \@sanitize@url \@href}%
	\providecommand \@href[1]{\@@startlink{#1}\@@href}%
	\providecommand \@@href[1]{\endgroup#1\@@endlink}%
	\providecommand \@sanitize@url [0]{\catcode `\\12\catcode `\$12\catcode
		`\&12\catcode `\#12\catcode `\^12\catcode `\_12\catcode `\%12\relax}%
	\providecommand \@@startlink[1]{}%
	\providecommand \@@endlink[0]{}%
	\providecommand \url  [0]{\begingroup\@sanitize@url \@url }%
	\providecommand \@url [1]{\endgroup\@href {#1}{\urlprefix }}%
	\providecommand \urlprefix  [0]{URL }%
	\providecommand \Eprint [0]{\href }%
	\providecommand \doibase [0]{http://dx.doi.org/}%
	\providecommand \selectlanguage [0]{\@gobble}%
	\providecommand \bibinfo  [0]{\@secondoftwo}%
	\providecommand \bibfield  [0]{\@secondoftwo}%
	\providecommand \translation [1]{[#1]}%
	\providecommand \BibitemOpen [0]{}%
	\providecommand \bibitemStop [0]{}%
	\providecommand \bibitemNoStop [0]{.\EOS\space}%
	\providecommand \EOS [0]{\spacefactor3000\relax}%
	\providecommand \BibitemShut  [1]{\csname bibitem#1\endcsname}%
	\let\auto@bib@innerbib\@empty
	\bibitem [{\citenamefont {Gisin}\ \emph {et~al.}(2002)\citenamefont {Gisin},
		\citenamefont {Ribordy}, \citenamefont {Tittel},\ and\ \citenamefont
		{Zbinden}}]{Gisin02}%
	\BibitemOpen
	\bibfield  {author} {\bibinfo {author} {\bibfnamefont {Nicolas}\ \bibnamefont
			{Gisin}}, \bibinfo {author} {\bibfnamefont {Gr\'egoire}\ \bibnamefont
			{Ribordy}}, \bibinfo {author} {\bibfnamefont {Wolfgang}\ \bibnamefont
			{Tittel}}, \ and\ \bibinfo {author} {\bibfnamefont {Hugo}\ \bibnamefont
			{Zbinden}},\ }\bibfield  {title} {\enquote {\bibinfo {title} {Quantum
				cryptography},}\ }\href@noop {} {\bibfield  {journal} {\bibinfo  {journal}
			{Rev. Mod. Phys.}\ }\textbf {\bibinfo {volume} {74}},\ \bibinfo {pages}
		{145--195} (\bibinfo {year} {2002})}\BibitemShut {NoStop}%
	\bibitem [{\citenamefont {O'brien}\ \emph {et~al.}(2009)\citenamefont
		{O'brien}, \citenamefont {Furusawa},\ and\ \citenamefont
		{Vu{\v{c}}kovi{\'c}}}]{OBrien09}%
	\BibitemOpen
	\bibfield  {author} {\bibinfo {author} {\bibfnamefont {Jeremy~L}\
			\bibnamefont {O'brien}}, \bibinfo {author} {\bibfnamefont {Akira}\
			\bibnamefont {Furusawa}}, \ and\ \bibinfo {author} {\bibfnamefont {Jelena}\
			\bibnamefont {Vu{\v{c}}kovi{\'c}}},\ }\bibfield  {title} {\enquote {\bibinfo
			{title} {Photonic quantum technologies},}\ }\href@noop {} {\bibfield
		{journal} {\bibinfo  {journal} {Nature Photonics}\ }\textbf {\bibinfo
			{volume} {3}},\ \bibinfo {pages} {687--695} (\bibinfo {year}
		{2009})}\BibitemShut {NoStop}%
	\bibitem [{\citenamefont {Shields}(2007)}]{Shields07}%
	\BibitemOpen
	\bibfield  {author} {\bibinfo {author} {\bibfnamefont {Andrew~J}\
			\bibnamefont {Shields}},\ }\bibfield  {title} {\enquote {\bibinfo {title}
			{Semiconductor quantum light sources},}\ }\href@noop {} {\bibfield  {journal}
		{\bibinfo  {journal} {Nature photonics}\ }\textbf {\bibinfo {volume} {1}},\
		\bibinfo {pages} {215--223} (\bibinfo {year} {2007})}\BibitemShut {NoStop}%
	\bibitem [{\citenamefont {Aharonovich}\ \emph {et~al.}(2016)\citenamefont
		{Aharonovich}, \citenamefont {Englund},\ and\ \citenamefont
		{Toth}}]{Aharonovich16}%
	\BibitemOpen
	\bibfield  {author} {\bibinfo {author} {\bibfnamefont {Igor}\ \bibnamefont
			{Aharonovich}}, \bibinfo {author} {\bibfnamefont {Dirk}\ \bibnamefont
			{Englund}}, \ and\ \bibinfo {author} {\bibfnamefont {Milos}\ \bibnamefont
			{Toth}},\ }\bibfield  {title} {\enquote {\bibinfo {title} {Solid-state
				single-photon emitters},}\ }\href@noop {} {\bibfield  {journal} {\bibinfo
			{journal} {Nature Photonics}\ }\textbf {\bibinfo {volume} {10}},\ \bibinfo
		{pages} {631--641} (\bibinfo {year} {2016})}\BibitemShut {NoStop}%
	\bibitem [{\citenamefont {Schr\"{o}der}\ \emph {et~al.}(2016)\citenamefont
		{Schr\"{o}der}, \citenamefont {Mouradian}, \citenamefont {Zheng},
		\citenamefont {Trusheim}, \citenamefont {Walsh}, \citenamefont {Chen},
		\citenamefont {Li}, \citenamefont {Bayn},\ and\ \citenamefont
		{Englund}}]{Schroeder16}%
	\BibitemOpen
	\bibfield  {author} {\bibinfo {author} {\bibfnamefont {Tim}\ \bibnamefont
			{Schr\"{o}der}}, \bibinfo {author} {\bibfnamefont {Sara~L.}\ \bibnamefont
			{Mouradian}}, \bibinfo {author} {\bibfnamefont {Jiabao}\ \bibnamefont
			{Zheng}}, \bibinfo {author} {\bibfnamefont {Matthew~E.}\ \bibnamefont
			{Trusheim}}, \bibinfo {author} {\bibfnamefont {Michael}\ \bibnamefont
			{Walsh}}, \bibinfo {author} {\bibfnamefont {Edward~H.}\ \bibnamefont {Chen}},
		\bibinfo {author} {\bibfnamefont {Luozhou}\ \bibnamefont {Li}}, \bibinfo
		{author} {\bibfnamefont {Igal}\ \bibnamefont {Bayn}}, \ and\ \bibinfo
		{author} {\bibfnamefont {Dirk}\ \bibnamefont {Englund}},\ }\bibfield  {title}
	{\enquote {\bibinfo {title} {Quantum nanophotonics in diamond},}\ }\href@noop
	{} {\bibfield  {journal} {\bibinfo  {journal} {J. Opt. Soc. Am. B}\ }\textbf
		{\bibinfo {volume} {33}},\ \bibinfo {pages} {B65--B83} (\bibinfo {year}
		{2016})}\BibitemShut {NoStop}%
	\bibitem [{\citenamefont {Michler}\ \emph {et~al.}(2000)\citenamefont
		{Michler}, \citenamefont {Kiraz}, \citenamefont {Becher}, \citenamefont
		{Schoenfeld}, \citenamefont {Petroff}, \citenamefont {Zhang}, \citenamefont
		{Hu},\ and\ \citenamefont {Imamoglu}}]{Michler00}%
	\BibitemOpen
	\bibfield  {author} {\bibinfo {author} {\bibfnamefont {P}~\bibnamefont
			{Michler}}, \bibinfo {author} {\bibfnamefont {A}~\bibnamefont {Kiraz}},
		\bibinfo {author} {\bibfnamefont {C}~\bibnamefont {Becher}}, \bibinfo
		{author} {\bibfnamefont {WV}~\bibnamefont {Schoenfeld}}, \bibinfo {author}
		{\bibfnamefont {PM}~\bibnamefont {Petroff}}, \bibinfo {author} {\bibfnamefont
			{Lidong}\ \bibnamefont {Zhang}}, \bibinfo {author} {\bibfnamefont
			{E}~\bibnamefont {Hu}}, \ and\ \bibinfo {author} {\bibfnamefont
			{A}~\bibnamefont {Imamoglu}},\ }\bibfield  {title} {\enquote {\bibinfo
			{title} {A quantum dot single-photon turnstile device},}\ }\href@noop {}
	{\bibfield  {journal} {\bibinfo  {journal} {Science}\ }\textbf {\bibinfo
			{volume} {290}},\ \bibinfo {pages} {2282--2285} (\bibinfo {year}
		{2000})}\BibitemShut {NoStop}%
	\bibitem [{\citenamefont {Strauf}\ \emph {et~al.}(2007)\citenamefont {Strauf},
		\citenamefont {Stoltz}, \citenamefont {Rakher}, \citenamefont {Coldren},
		\citenamefont {Petroff},\ and\ \citenamefont {Bouwmeester}}]{Strauf07}%
	\BibitemOpen
	\bibfield  {author} {\bibinfo {author} {\bibfnamefont {Stefan}\ \bibnamefont
			{Strauf}}, \bibinfo {author} {\bibfnamefont {Nick~G}\ \bibnamefont {Stoltz}},
		\bibinfo {author} {\bibfnamefont {Matthew~T}\ \bibnamefont {Rakher}},
		\bibinfo {author} {\bibfnamefont {Larry~A}\ \bibnamefont {Coldren}}, \bibinfo
		{author} {\bibfnamefont {Pierre~M}\ \bibnamefont {Petroff}}, \ and\ \bibinfo
		{author} {\bibfnamefont {Dirk}\ \bibnamefont {Bouwmeester}},\ }\bibfield
	{title} {\enquote {\bibinfo {title} {High-frequency single-photon source with
				polarization control},}\ }\href@noop {} {\bibfield  {journal} {\bibinfo
			{journal} {Nature photonics}\ }\textbf {\bibinfo {volume} {1}},\ \bibinfo
		{pages} {704--708} (\bibinfo {year} {2007})}\BibitemShut {NoStop}%
	\bibitem [{\citenamefont {Englund}\ \emph {et~al.}(2010)\citenamefont
		{Englund}, \citenamefont {Shields}, \citenamefont {Rivoire}, \citenamefont
		{Hatami}, \citenamefont {Vu\v{c}kovi\'c}, \citenamefont {Park},\ and\
		\citenamefont {Lukin}}]{Englund10}%
	\BibitemOpen
	\bibfield  {author} {\bibinfo {author} {\bibfnamefont {D.}~\bibnamefont
			{Englund}}, \bibinfo {author} {\bibfnamefont {B.}~\bibnamefont {Shields}},
		\bibinfo {author} {\bibfnamefont {K.}~\bibnamefont {Rivoire}}, \bibinfo
		{author} {\bibfnamefont {F.}~\bibnamefont {Hatami}}, \bibinfo {author}
		{\bibfnamefont {J.}~\bibnamefont {Vu\v{c}kovi\'c}}, \bibinfo {author}
		{\bibfnamefont {H.}~\bibnamefont {Park}}, \ and\ \bibinfo {author}
		{\bibfnamefont {M.~D.}\ \bibnamefont {Lukin}},\ }\bibfield  {title} {\enquote
		{\bibinfo {title} {Deterministic coupling of a single nitrogen vacancy center
				to a photonic crystal cavity},}\ }\href@noop {} {\bibfield  {journal}
		{\bibinfo  {journal} {Nano Letters}\ }\textbf {\bibinfo {volume} {10}},\
		\bibinfo {pages} {3922} (\bibinfo {year} {2010})}\BibitemShut {NoStop}%
	\bibitem [{\citenamefont {Gazzano}\ \emph {et~al.}(2013)\citenamefont
		{Gazzano}, \citenamefont {de~Vasconcellos}, \citenamefont {Arnold},
		\citenamefont {Nowak}, \citenamefont {Galopin}, \citenamefont {Sagnes},
		\citenamefont {Lanco}, \citenamefont {Lemaitre},\ and\ \citenamefont
		{Senellart}}]{Gazzano13}%
	\BibitemOpen
	\bibfield  {author} {\bibinfo {author} {\bibfnamefont {O.}~\bibnamefont
			{Gazzano}}, \bibinfo {author} {\bibfnamefont {S.~Michaelis}\ \bibnamefont
			{de~Vasconcellos}}, \bibinfo {author} {\bibfnamefont {C.}~\bibnamefont
			{Arnold}}, \bibinfo {author} {\bibfnamefont {A.}~\bibnamefont {Nowak}},
		\bibinfo {author} {\bibfnamefont {E.}~\bibnamefont {Galopin}}, \bibinfo
		{author} {\bibfnamefont {I.}~\bibnamefont {Sagnes}}, \bibinfo {author}
		{\bibfnamefont {L.}~\bibnamefont {Lanco}}, \bibinfo {author} {\bibfnamefont
			{A.}~\bibnamefont {Lemaitre}}, \ and\ \bibinfo {author} {\bibfnamefont
			{P.}~\bibnamefont {Senellart}},\ }\bibfield  {title} {\enquote {\bibinfo
			{title} {Bright solid-state sources of indistinguishable single photons},}\
	}\href@noop {} {\bibfield  {journal} {\bibinfo  {journal} {Nature
			Communications}\ }\textbf {\bibinfo {volume} {4}},\ \bibinfo {pages} {1425}
	(\bibinfo {year} {2013})}\BibitemShut {NoStop}%
\bibitem [{\citenamefont {Albrecht}\ \emph {et~al.}(2013)\citenamefont
	{Albrecht}, \citenamefont {Bommer}, \citenamefont {Deutsch}, \citenamefont
	{Reichel},\ and\ \citenamefont {Becher}}]{Albrecht13}%
\BibitemOpen
\bibfield  {author} {\bibinfo {author} {\bibfnamefont {Roland}\ \bibnamefont
		{Albrecht}}, \bibinfo {author} {\bibfnamefont {Alexander}\ \bibnamefont
		{Bommer}}, \bibinfo {author} {\bibfnamefont {Christian}\ \bibnamefont
		{Deutsch}}, \bibinfo {author} {\bibfnamefont {Jakob}\ \bibnamefont
		{Reichel}}, \ and\ \bibinfo {author} {\bibfnamefont {Christoph}\ \bibnamefont
		{Becher}},\ }\bibfield  {title} {\enquote {\bibinfo {title} {Coupling of a
			single nitrogen-vacancy center in diamond to a fiber-based microcavity},}\
}\href@noop {} {\bibfield  {journal} {\bibinfo  {journal} {Physical review
		letters}\ }\textbf {\bibinfo {volume} {110}},\ \bibinfo {pages} {243602}
(\bibinfo {year} {2013})}\BibitemShut {NoStop}%
\bibitem [{\citenamefont {Kaupp}\ \emph {et~al.}(2016)\citenamefont {Kaupp},
	\citenamefont {H\"ummer}, \citenamefont {Mader}, \citenamefont {Schlederer},
	\citenamefont {Benedikter}, \citenamefont {Haeusser}, \citenamefont {Chang},
	\citenamefont {Fedder}, \citenamefont {H\"ansch},\ and\ \citenamefont
	{Hunger}}]{Kaupp16}%
\BibitemOpen
\bibfield  {author} {\bibinfo {author} {\bibfnamefont {Hanno}\ \bibnamefont
		{Kaupp}}, \bibinfo {author} {\bibfnamefont {Thomas}\ \bibnamefont
		{H\"ummer}}, \bibinfo {author} {\bibfnamefont {Matthias}\ \bibnamefont
		{Mader}}, \bibinfo {author} {\bibfnamefont {Benedikt}\ \bibnamefont
		{Schlederer}}, \bibinfo {author} {\bibfnamefont {Julia}\ \bibnamefont
		{Benedikter}}, \bibinfo {author} {\bibfnamefont {Philip}\ \bibnamefont
		{Haeusser}}, \bibinfo {author} {\bibfnamefont {Huan-Cheng}\ \bibnamefont
		{Chang}}, \bibinfo {author} {\bibfnamefont {Helmut}\ \bibnamefont {Fedder}},
	\bibinfo {author} {\bibfnamefont {Theodor~W.}\ \bibnamefont {H\"ansch}}, \
	and\ \bibinfo {author} {\bibfnamefont {David}\ \bibnamefont {Hunger}},\
}\bibfield  {title} {\enquote {\bibinfo {title} {Purcell-enhanced
		single-photon emission from nitrogen-vacancy centers coupled to a tunable
		microcavity},}\ }\href@noop {} {\bibfield  {journal} {\bibinfo  {journal}
	{Phys. Rev. Applied}\ }\textbf {\bibinfo {volume} {6}},\ \bibinfo {pages}
{054010} (\bibinfo {year} {2016})}\BibitemShut {NoStop}%
\bibitem [{\citenamefont {Laucht}\ \emph {et~al.}(2012)\citenamefont {Laucht},
	\citenamefont {P\"utz}, \citenamefont {G\"unthner}, \citenamefont {Hauke},
	\citenamefont {Saive}, \citenamefont {Fr\'ed\'erick}, \citenamefont
	{Bichler}, \citenamefont {Amann}, \citenamefont {Holleitner}, \citenamefont
	{Kaniber},\ and\ \citenamefont {Finley}}]{Laucht12}%
\BibitemOpen
\bibfield  {author} {\bibinfo {author} {\bibfnamefont {A.}~\bibnamefont
		{Laucht}}, \bibinfo {author} {\bibfnamefont {S.}~\bibnamefont {P\"utz}},
	\bibinfo {author} {\bibfnamefont {T.}~\bibnamefont {G\"unthner}}, \bibinfo
	{author} {\bibfnamefont {N.}~\bibnamefont {Hauke}}, \bibinfo {author}
	{\bibfnamefont {R.}~\bibnamefont {Saive}}, \bibinfo {author} {\bibfnamefont
		{S.}~\bibnamefont {Fr\'ed\'erick}}, \bibinfo {author} {\bibfnamefont
		{M.}~\bibnamefont {Bichler}}, \bibinfo {author} {\bibfnamefont {M.-C.}\
		\bibnamefont {Amann}}, \bibinfo {author} {\bibfnamefont {A.~W.}\ \bibnamefont
		{Holleitner}}, \bibinfo {author} {\bibfnamefont {M.}~\bibnamefont {Kaniber}},
	\ and\ \bibinfo {author} {\bibfnamefont {J.~J.}\ \bibnamefont {Finley}},\
}\bibfield  {title} {\enquote {\bibinfo {title} {A waveguide-coupled on-chip
		single-photon source},}\ }\href@noop {} {\bibfield  {journal} {\bibinfo
	{journal} {Phys. Rev. X}\ }\textbf {\bibinfo {volume} {2}},\ \bibinfo {pages}
{011014} (\bibinfo {year} {2012})}\BibitemShut {NoStop}%
\bibitem [{\citenamefont {Liebermeister}\ \emph {et~al.}(2014)\citenamefont
	{Liebermeister}, \citenamefont {Petersen}, \citenamefont {Münchow},
	\citenamefont {Burchardt}, \citenamefont {Hermelbracht}, \citenamefont
	{Tashima}, \citenamefont {Schell}, \citenamefont {Benson}, \citenamefont
	{Meinhardt}, \citenamefont {Krueger}, \citenamefont {Stiebeiner},
	\citenamefont {Rauschenbeutel}, \citenamefont {Weinfurter},\ and\
	\citenamefont {Weber}}]{Liebermeister14}%
\BibitemOpen
\bibfield  {author} {\bibinfo {author} {\bibfnamefont {Lars}\ \bibnamefont
		{Liebermeister}}, \bibinfo {author} {\bibfnamefont {Fabian}\ \bibnamefont
		{Petersen}}, \bibinfo {author} {\bibfnamefont {Asmus~v.}\ \bibnamefont
		{Münchow}}, \bibinfo {author} {\bibfnamefont {Daniel}\ \bibnamefont
		{Burchardt}}, \bibinfo {author} {\bibfnamefont {Juliane}\ \bibnamefont
		{Hermelbracht}}, \bibinfo {author} {\bibfnamefont {Toshiyuki}\ \bibnamefont
		{Tashima}}, \bibinfo {author} {\bibfnamefont {Andreas~W.}\ \bibnamefont
		{Schell}}, \bibinfo {author} {\bibfnamefont {Oliver}\ \bibnamefont {Benson}},
	\bibinfo {author} {\bibfnamefont {Thomas}\ \bibnamefont {Meinhardt}},
	\bibinfo {author} {\bibfnamefont {Anke}\ \bibnamefont {Krueger}}, \bibinfo
	{author} {\bibfnamefont {Ariane}\ \bibnamefont {Stiebeiner}}, \bibinfo
	{author} {\bibfnamefont {Arno}\ \bibnamefont {Rauschenbeutel}}, \bibinfo
	{author} {\bibfnamefont {Harald}\ \bibnamefont {Weinfurter}}, \ and\ \bibinfo
	{author} {\bibfnamefont {Markus}\ \bibnamefont {Weber}},\ }\bibfield  {title}
{\enquote {\bibinfo {title} {Tapered fiber coupling of single photons emitted
			by a deterministically positioned single nitrogen vacancy center},}\
}\href@noop {} {\bibfield  {journal} {\bibinfo  {journal} {Applied Physics
		Letters}\ }\textbf {\bibinfo {volume} {104}} (\bibinfo {year}
{2014})}\BibitemShut {NoStop}%
\bibitem [{\citenamefont {Chu}\ \emph {et~al.}(2014)\citenamefont {Chu},
	\citenamefont {Brenner}, \citenamefont {Chen}, \citenamefont {Ghosh},
	\citenamefont {Hollingsworth}, \citenamefont {Sandoghdar},\ and\
	\citenamefont {G\"{o}tzinger}}]{Chu14}%
\BibitemOpen
\bibfield  {author} {\bibinfo {author} {\bibfnamefont {X.-L.}\ \bibnamefont
		{Chu}}, \bibinfo {author} {\bibfnamefont {T.~J.~K.}\ \bibnamefont {Brenner}},
	\bibinfo {author} {\bibfnamefont {X.-W.}\ \bibnamefont {Chen}}, \bibinfo
	{author} {\bibfnamefont {Y.}~\bibnamefont {Ghosh}}, \bibinfo {author}
	{\bibfnamefont {J.~A.}\ \bibnamefont {Hollingsworth}}, \bibinfo {author}
	{\bibfnamefont {V.}~\bibnamefont {Sandoghdar}}, \ and\ \bibinfo {author}
	{\bibfnamefont {S.}~\bibnamefont {G\"{o}tzinger}},\ }\bibfield  {title}
{\enquote {\bibinfo {title} {Experimental realization of an optical antenna
			designed for collecting 99\% of photons from a quantum emitter},}\
}\href@noop {} {\bibfield  {journal} {\bibinfo  {journal} {Optica}\ }\textbf
{\bibinfo {volume} {1}},\ \bibinfo {pages} {203--208} (\bibinfo {year}
{2014})}\BibitemShut {NoStop}%
\bibitem [{\citenamefont {Li}\ \emph {et~al.}(2015)\citenamefont {Li},
	\citenamefont {Chen}, \citenamefont {Zheng}, \citenamefont {Mouradian},
	\citenamefont {Dolde}, \citenamefont {Schröder}, \citenamefont {Karaveli},
	\citenamefont {Markham}, \citenamefont {Twitchen},\ and\ \citenamefont
	{Englund}}]{Li15}%
\BibitemOpen
\bibfield  {author} {\bibinfo {author} {\bibfnamefont {Luozhou}\ \bibnamefont
		{Li}}, \bibinfo {author} {\bibfnamefont {Edward~H.}\ \bibnamefont {Chen}},
	\bibinfo {author} {\bibfnamefont {Jiabao}\ \bibnamefont {Zheng}}, \bibinfo
	{author} {\bibfnamefont {Sara~L.}\ \bibnamefont {Mouradian}}, \bibinfo
	{author} {\bibfnamefont {Florian}\ \bibnamefont {Dolde}}, \bibinfo {author}
	{\bibfnamefont {Tim}\ \bibnamefont {Schröder}}, \bibinfo {author}
	{\bibfnamefont {Sinan}\ \bibnamefont {Karaveli}}, \bibinfo {author}
	{\bibfnamefont {Matthew~L.}\ \bibnamefont {Markham}}, \bibinfo {author}
	{\bibfnamefont {Daniel~J.}\ \bibnamefont {Twitchen}}, \ and\ \bibinfo
	{author} {\bibfnamefont {Dirk}\ \bibnamefont {Englund}},\ }\bibfield  {title}
{\enquote {\bibinfo {title} {Efficient photon collection from a nitrogen
			vacancy center in a circular bullseye grating},}\ }\href@noop {} {\bibfield
	{journal} {\bibinfo  {journal} {Nano Letters}\ }\textbf {\bibinfo {volume}
		{15}},\ \bibinfo {pages} {1493--1497} (\bibinfo {year} {2015})}\BibitemShut
{NoStop}%
\bibitem [{\citenamefont {Schlehahn}\ \emph {et~al.}(2016)\citenamefont
	{Schlehahn}, \citenamefont {Thoma}, \citenamefont {Munnelly}, \citenamefont
	{Kamp}, \citenamefont {H{\"o}fling}, \citenamefont {Heindel}, \citenamefont
	{Schneider},\ and\ \citenamefont {Reitzenstein}}]{Schlehahn16}%
\BibitemOpen
\bibfield  {author} {\bibinfo {author} {\bibfnamefont {A}~\bibnamefont
		{Schlehahn}}, \bibinfo {author} {\bibfnamefont {A}~\bibnamefont {Thoma}},
	\bibinfo {author} {\bibfnamefont {P}~\bibnamefont {Munnelly}}, \bibinfo
	{author} {\bibfnamefont {M}~\bibnamefont {Kamp}}, \bibinfo {author}
	{\bibfnamefont {Sven}\ \bibnamefont {H{\"o}fling}}, \bibinfo {author}
	{\bibfnamefont {T}~\bibnamefont {Heindel}}, \bibinfo {author} {\bibfnamefont
		{C}~\bibnamefont {Schneider}}, \ and\ \bibinfo {author} {\bibfnamefont
		{S}~\bibnamefont {Reitzenstein}},\ }\bibfield  {title} {\enquote {\bibinfo
		{title} {An electrically driven cavity-enhanced source of indistinguishable
			photons with 61\% overall efficiency},}\ }\href@noop {} {\bibfield  {journal}
	{\bibinfo  {journal} {APL Photonics}\ }\textbf {\bibinfo {volume} {1}},\
	\bibinfo {pages} {011301} (\bibinfo {year} {2016})}\BibitemShut {NoStop}%
\bibitem [{\citenamefont {Steiner}\ \emph {et~al.}(2007)\citenamefont
	{Steiner}, \citenamefont {Hartschuh}, \citenamefont {Korlacki},\ and\
	\citenamefont {Meixner}}]{Steiner07}%
\BibitemOpen
\bibfield  {author} {\bibinfo {author} {\bibfnamefont {Mathias}\ \bibnamefont
		{Steiner}}, \bibinfo {author} {\bibfnamefont {Achim}\ \bibnamefont
		{Hartschuh}}, \bibinfo {author} {\bibfnamefont {Rafał}\ \bibnamefont
		{Korlacki}}, \ and\ \bibinfo {author} {\bibfnamefont {Alfred~J.}\
		\bibnamefont {Meixner}},\ }\bibfield  {title} {\enquote {\bibinfo {title}
		{Highly efficient, tunable single photon source based on single molecules},}\
}\href@noop {} {\bibfield  {journal} {\bibinfo  {journal} {Applied Physics
		Letters}\ }\textbf {\bibinfo {volume} {90}} (\bibinfo {year}
{2007})}\BibitemShut {NoStop}%
\bibitem [{\citenamefont {Toninelli}\ \emph {et~al.}(2010)\citenamefont
	{Toninelli}, \citenamefont {Delley}, \citenamefont {St{\"o}ferle},
	\citenamefont {Renn}, \citenamefont {G{\"o}tzinger},\ and\ \citenamefont
	{Sandoghdar}}]{Toninelli10}%
\BibitemOpen
\bibfield  {author} {\bibinfo {author} {\bibfnamefont {C.}~\bibnamefont
		{Toninelli}}, \bibinfo {author} {\bibfnamefont {Y.}~\bibnamefont {Delley}},
	\bibinfo {author} {\bibfnamefont {T.}~\bibnamefont {St{\"o}ferle}}, \bibinfo
	{author} {\bibfnamefont {A.}~\bibnamefont {Renn}}, \bibinfo {author}
	{\bibfnamefont {S.}~\bibnamefont {G{\"o}tzinger}}, \ and\ \bibinfo {author}
	{\bibfnamefont {V.}~\bibnamefont {Sandoghdar}},\ }\bibfield  {title}
{\enquote {\bibinfo {title} {A scanning microcavity for in situ control of
			single-molecule emission},}\ }\href@noop {} {\bibfield  {journal} {\bibinfo
		{journal} {Appl. Phys. Lett.}\ }\textbf {\bibinfo {volume} {97}},\ \bibinfo
	{pages} {021107} (\bibinfo {year} {2010})}\BibitemShut {NoStop}%
\bibitem [{\citenamefont {Miura}\ \emph {et~al.}(2014)\citenamefont {Miura},
	\citenamefont {Imamura}, \citenamefont {Ohta}, \citenamefont {Ishii},
	\citenamefont {Liu}, \citenamefont {Shimada}, \citenamefont {Iwamoto},
	\citenamefont {Arakawa},\ and\ \citenamefont {Kato}}]{Miura14}%
\BibitemOpen
\bibfield  {author} {\bibinfo {author} {\bibfnamefont {R.}~\bibnamefont
		{Miura}}, \bibinfo {author} {\bibfnamefont {S.}~\bibnamefont {Imamura}},
	\bibinfo {author} {\bibfnamefont {R.}~\bibnamefont {Ohta}}, \bibinfo {author}
	{\bibfnamefont {A.}~\bibnamefont {Ishii}}, \bibinfo {author} {\bibfnamefont
		{X.}~\bibnamefont {Liu}}, \bibinfo {author} {\bibfnamefont {T.}~\bibnamefont
		{Shimada}}, \bibinfo {author} {\bibfnamefont {S.}~\bibnamefont {Iwamoto}},
	\bibinfo {author} {\bibfnamefont {Y.}~\bibnamefont {Arakawa}}, \ and\
	\bibinfo {author} {\bibfnamefont {Y.~K.}\ \bibnamefont {Kato}},\ }\bibfield
{title} {\enquote {\bibinfo {title} {Ultralow mode-volume photonic crystal
			nanobeam cavities for high-efficiency coupling to individual carbon nanotube
			emitters},}\ }\href@noop {} {\bibfield  {journal} {\bibinfo  {journal}
		{Nature Communications}\ }\textbf {\bibinfo {volume} {5}},\ \bibinfo {pages}
	{5580} (\bibinfo {year} {2014})}\BibitemShut {NoStop}%
\bibitem [{\citenamefont {Pyatkov}\ \emph {et~al.}(2016)\citenamefont
	{Pyatkov}, \citenamefont {F{\"u}tterling}, \citenamefont {Khasminskaya},
	\citenamefont {Flavel}, \citenamefont {Hennrich}, \citenamefont {Kappes},
	\citenamefont {Krupke},\ and\ \citenamefont {Pernice}}]{Pyatkov16}%
\BibitemOpen
\bibfield  {author} {\bibinfo {author} {\bibfnamefont {F.}~\bibnamefont
		{Pyatkov}}, \bibinfo {author} {\bibfnamefont {V.}~\bibnamefont
		{F{\"u}tterling}}, \bibinfo {author} {\bibfnamefont {S.}~\bibnamefont
		{Khasminskaya}}, \bibinfo {author} {\bibfnamefont {B.~S.}\ \bibnamefont
		{Flavel}}, \bibinfo {author} {\bibfnamefont {F.}~\bibnamefont {Hennrich}},
	\bibinfo {author} {\bibfnamefont {M.~M.}\ \bibnamefont {Kappes}}, \bibinfo
	{author} {\bibfnamefont {R.}~\bibnamefont {Krupke}}, \ and\ \bibinfo {author}
	{\bibfnamefont {W.~H.~P.}\ \bibnamefont {Pernice}},\ }\bibfield  {title}
{\enquote {\bibinfo {title} {Cavity-enhanced light emission from electrically
			driven carbon nanotubes},}\ }\href@noop {} {\bibfield  {journal} {\bibinfo
		{journal} {Nature Photonics}\ }\textbf {\bibinfo {volume} {10}},\ \bibinfo
	{pages} {420} (\bibinfo {year} {2016})}\BibitemShut {NoStop}%
\bibitem [{\citenamefont {Jeantet}\ \emph {et~al.}(2016)\citenamefont
	{Jeantet}, \citenamefont {Chassagneux}, \citenamefont {Raynaud},
	\citenamefont {Roussignol}, \citenamefont {Lauret}, \citenamefont {Besga},
	\citenamefont {Esteve}, \citenamefont {Reichel},\ and\ \citenamefont
	{Voisin}}]{Jeantet16}%
\BibitemOpen
\bibfield  {author} {\bibinfo {author} {\bibfnamefont {A.}~\bibnamefont
		{Jeantet}}, \bibinfo {author} {\bibfnamefont {Y}~\bibnamefont {Chassagneux}},
	\bibinfo {author} {\bibfnamefont {C.}~\bibnamefont {Raynaud}}, \bibinfo
	{author} {\bibfnamefont {P.}~\bibnamefont {Roussignol}}, \bibinfo {author}
	{\bibfnamefont {J.~S.}\ \bibnamefont {Lauret}}, \bibinfo {author}
	{\bibfnamefont {B.}~\bibnamefont {Besga}}, \bibinfo {author} {\bibfnamefont
		{J.}~\bibnamefont {Esteve}}, \bibinfo {author} {\bibfnamefont
		{J.}~\bibnamefont {Reichel}}, \ and\ \bibinfo {author} {\bibfnamefont
		{C.}~\bibnamefont {Voisin}},\ }\bibfield  {title} {\enquote {\bibinfo {title}
		{Widely tunable single-photon source from a carbon nanotube in the purcell
			regime},}\ }\href@noop {} {\bibfield  {journal} {\bibinfo  {journal} {Phys.
			Rev. Lett.}\ }\textbf {\bibinfo {volume} {116}},\ \bibinfo {pages} {247402}
	(\bibinfo {year} {2016})}\BibitemShut {NoStop}%
\bibitem [{\citenamefont {Faraon}\ \emph {et~al.}(2012)\citenamefont {Faraon},
	\citenamefont {Santori}, \citenamefont {Huang}, \citenamefont {Acosta},\ and\
	\citenamefont {Beausoleil}}]{Faraon12}%
\BibitemOpen
\bibfield  {author} {\bibinfo {author} {\bibfnamefont {A.}~\bibnamefont
		{Faraon}}, \bibinfo {author} {\bibfnamefont {C.}~\bibnamefont {Santori}},
	\bibinfo {author} {\bibfnamefont {Z.}~\bibnamefont {Huang}}, \bibinfo
	{author} {\bibfnamefont {V.~M.}\ \bibnamefont {Acosta}}, \ and\ \bibinfo
	{author} {\bibfnamefont {R.~G.}\ \bibnamefont {Beausoleil}},\ }\bibfield
{title} {\enquote {\bibinfo {title} {Coupling of nitrogen-vacancy centers to
			photonic crystal cavities in monocrystalline diamond},}\ }\href@noop {}
{\bibfield  {journal} {\bibinfo  {journal} {Phys. Rev. Lett.}\ }\textbf
	{\bibinfo {volume} {109}},\ \bibinfo {pages} {033604} (\bibinfo {year}
	{2012})}\BibitemShut {NoStop}%
\bibitem [{\citenamefont {Johnson}\ \emph {et~al.}(2015)\citenamefont
	{Johnson}, \citenamefont {Dolan}, \citenamefont {Grange}, \citenamefont
	{Trichet}, \citenamefont {Hornecker}, \citenamefont {Chen}, \citenamefont
	{Weng}, \citenamefont {Hughes}, \citenamefont {Watt}, \citenamefont
	{Auff\`{e}ves},\ and\ \citenamefont {Smith}}]{Johnson15}%
\BibitemOpen
\bibfield  {author} {\bibinfo {author} {\bibfnamefont {S}~\bibnamefont
		{Johnson}}, \bibinfo {author} {\bibfnamefont {P~R}\ \bibnamefont {Dolan}},
	\bibinfo {author} {\bibfnamefont {T}~\bibnamefont {Grange}}, \bibinfo
	{author} {\bibfnamefont {A~A~P}\ \bibnamefont {Trichet}}, \bibinfo {author}
	{\bibfnamefont {G}~\bibnamefont {Hornecker}}, \bibinfo {author}
	{\bibfnamefont {Y~C}\ \bibnamefont {Chen}}, \bibinfo {author} {\bibfnamefont
		{L}~\bibnamefont {Weng}}, \bibinfo {author} {\bibfnamefont {G~M}\
		\bibnamefont {Hughes}}, \bibinfo {author} {\bibfnamefont {A~A~R}\
		\bibnamefont {Watt}}, \bibinfo {author} {\bibfnamefont {A}~\bibnamefont
		{Auff\`{e}ves}}, \ and\ \bibinfo {author} {\bibfnamefont {J~M}\ \bibnamefont
		{Smith}},\ }\bibfield  {title} {\enquote {\bibinfo {title} {Tunable cavity
			coupling of the zero phonon line of a nitrogen-vacancy defect in diamond},}\
}\href@noop {} {\bibfield  {journal} {\bibinfo  {journal} {New Journal of
		Physics}\ }\textbf {\bibinfo {volume} {17}},\ \bibinfo {pages} {122003}
(\bibinfo {year} {2015})}\BibitemShut {NoStop}%
\bibitem [{\citenamefont {Lee}\ \emph {et~al.}(2012)\citenamefont {Lee},
	\citenamefont {Aharonovich}, \citenamefont {Magyar}, \citenamefont {Rol},\
	and\ \citenamefont {Hu}}]{Lee12}%
\BibitemOpen
\bibfield  {author} {\bibinfo {author} {\bibfnamefont {J.~C.}\ \bibnamefont
		{Lee}}, \bibinfo {author} {\bibfnamefont {I.}~\bibnamefont {Aharonovich}},
	\bibinfo {author} {\bibfnamefont {A.~P.}\ \bibnamefont {Magyar}}, \bibinfo
	{author} {\bibfnamefont {F.}~\bibnamefont {Rol}}, \ and\ \bibinfo {author}
	{\bibfnamefont {E.~L.}\ \bibnamefont {Hu}},\ }\bibfield  {title} {\enquote
	{\bibinfo {title} {Coupling of silicon-vacancy centers to a single crystal
			diamond cavity},}\ }\href@noop {} {\bibfield  {journal} {\bibinfo  {journal}
		{Optics Express}\ }\textbf {\bibinfo {volume} {20}},\ \bibinfo {pages} {8891}
	(\bibinfo {year} {2012})}\BibitemShut {NoStop}%
\bibitem [{\citenamefont {Riedrich-M{\"o}ller}\ \emph
	{et~al.}(2012)\citenamefont {Riedrich-M{\"o}ller}, \citenamefont {Kipfstuhl},
	\citenamefont {Hepp}, \citenamefont {Neu}, \citenamefont {Pauly},
	\citenamefont {M{\"u}cklich}, \citenamefont {Baur}, \citenamefont {Wandt},
	\citenamefont {Wolff}, \citenamefont {Fischer}, \citenamefont {Gsell},
	\citenamefont {Schreck},\ and\ \citenamefont {Becher}}]{Riedrich12}%
\BibitemOpen
\bibfield  {author} {\bibinfo {author} {\bibfnamefont {J.}~\bibnamefont
		{Riedrich-M{\"o}ller}}, \bibinfo {author} {\bibfnamefont {L.}~\bibnamefont
		{Kipfstuhl}}, \bibinfo {author} {\bibfnamefont {C.}~\bibnamefont {Hepp}},
	\bibinfo {author} {\bibfnamefont {E.}~\bibnamefont {Neu}}, \bibinfo {author}
	{\bibfnamefont {C.}~\bibnamefont {Pauly}}, \bibinfo {author} {\bibfnamefont
		{F.}~\bibnamefont {M{\"u}cklich}}, \bibinfo {author} {\bibfnamefont
		{A.}~\bibnamefont {Baur}}, \bibinfo {author} {\bibfnamefont {M.}~\bibnamefont
		{Wandt}}, \bibinfo {author} {\bibfnamefont {S.}~\bibnamefont {Wolff}},
	\bibinfo {author} {\bibfnamefont {M.}~\bibnamefont {Fischer}}, \bibinfo
	{author} {\bibfnamefont {S.}~\bibnamefont {Gsell}}, \bibinfo {author}
	{\bibfnamefont {M.}~\bibnamefont {Schreck}}, \ and\ \bibinfo {author}
	{\bibfnamefont {C.}~\bibnamefont {Becher}},\ }\bibfield  {title} {\enquote
	{\bibinfo {title} {One- and two-dimensional photonic crystal microcavities in
			single crystal diamond},}\ }\href@noop {} {\bibfield  {journal} {\bibinfo
		{journal} {Nature Nanotechnology}\ }\textbf {\bibinfo {volume} {7}},\
	\bibinfo {pages} {69} (\bibinfo {year} {2012})}\BibitemShut {NoStop}%
\bibitem [{\citenamefont {Riedrich-M\"oller}\ \emph {et~al.}(2014)\citenamefont
	{Riedrich-M\"oller}, \citenamefont {Arend}, \citenamefont {Pauly},
	\citenamefont {M\"ucklich}, \citenamefont {Fischer}, \citenamefont {Gsell},
	\citenamefont {Schreck},\ and\ \citenamefont {Becher}}]{Riedrich14}%
\BibitemOpen
\bibfield  {author} {\bibinfo {author} {\bibfnamefont {Janine}\ \bibnamefont
		{Riedrich-M\"oller}}, \bibinfo {author} {\bibfnamefont {Carsten}\
		\bibnamefont {Arend}}, \bibinfo {author} {\bibfnamefont {Christoph}\
		\bibnamefont {Pauly}}, \bibinfo {author} {\bibfnamefont {Frank}\ \bibnamefont
		{M\"ucklich}}, \bibinfo {author} {\bibfnamefont {Martin}\ \bibnamefont
		{Fischer}}, \bibinfo {author} {\bibfnamefont {Stefan}\ \bibnamefont {Gsell}},
	\bibinfo {author} {\bibfnamefont {Matthias}\ \bibnamefont {Schreck}}, \ and\
	\bibinfo {author} {\bibfnamefont {Christoph}\ \bibnamefont {Becher}},\
}\bibfield  {title} {\enquote {\bibinfo {title} {Deterministic coupling of a
		single silicon-vacancy color center to a photonic crystal cavity in
		diamond},}\ }\href@noop {} {\bibfield  {journal} {\bibinfo  {journal} {Nano
		Letters}\ }\textbf {\bibinfo {volume} {14}},\ \bibinfo {pages} {5281--5287}
(\bibinfo {year} {2014})}\BibitemShut {NoStop}%
\bibitem [{\citenamefont {Sipahigil}\ \emph {et~al.}(2016)\citenamefont
	{Sipahigil}, \citenamefont {Evans}, \citenamefont {Sukachev}, \citenamefont
	{Burek}, \citenamefont {Borregaard}, \citenamefont {Bhaskar}, \citenamefont
	{Nguyen}, \citenamefont {Pacheco}, \citenamefont {Atikian}, \citenamefont
	{Meuwly} \emph {et~al.}}]{Sipahigil16}%
\BibitemOpen
\bibfield  {author} {\bibinfo {author} {\bibfnamefont {A}~\bibnamefont
		{Sipahigil}}, \bibinfo {author} {\bibfnamefont {RE}~\bibnamefont {Evans}},
	\bibinfo {author} {\bibfnamefont {DD}~\bibnamefont {Sukachev}}, \bibinfo
	{author} {\bibfnamefont {MJ}~\bibnamefont {Burek}}, \bibinfo {author}
	{\bibfnamefont {J}~\bibnamefont {Borregaard}}, \bibinfo {author}
	{\bibfnamefont {MK}~\bibnamefont {Bhaskar}}, \bibinfo {author} {\bibfnamefont
		{CT}~\bibnamefont {Nguyen}}, \bibinfo {author} {\bibfnamefont
		{JL}~\bibnamefont {Pacheco}}, \bibinfo {author} {\bibfnamefont
		{HA}~\bibnamefont {Atikian}}, \bibinfo {author} {\bibfnamefont
		{C}~\bibnamefont {Meuwly}},  \emph {et~al.},\ }\bibfield  {title} {\enquote
	{\bibinfo {title} {An integrated diamond nanophotonics platform for
			quantum-optical networks},}\ }\href@noop {} {\bibfield  {journal} {\bibinfo
		{journal} {Science}\ }\textbf {\bibinfo {volume} {354}},\ \bibinfo {pages}
	{847--850} (\bibinfo {year} {2016})}\BibitemShut {NoStop}%
\bibitem [{\citenamefont {Vahala}(2003)}]{Vahala03}%
\BibitemOpen
\bibfield  {author} {\bibinfo {author} {\bibfnamefont {Kerry~J.}\
		\bibnamefont {Vahala}},\ }\bibfield  {title} {\enquote {\bibinfo {title}
		{Optical microcavities},}\ }\href {\doibase 10.1038/nature01939} {\bibfield
	{journal} {\bibinfo  {journal} {Nature}\ }\textbf {\bibinfo {volume} {424}},\
	\bibinfo {pages} {839--846} (\bibinfo {year} {2003})}\BibitemShut {NoStop}%
\bibitem [{\citenamefont {Auff{\`e}ves}\ \emph {et~al.}(2010)\citenamefont
	{Auff{\`e}ves}, \citenamefont {Gerace}, \citenamefont {G{\'e}rard},
	\citenamefont {{Santos}}, \citenamefont {Andreani},\ and\
	\citenamefont {Poizat}}]{Auffeves10}%
\BibitemOpen
\bibfield  {author} {\bibinfo {author} {\bibfnamefont {A.}~\bibnamefont
		{Auff{\`e}ves}}, \bibinfo {author} {\bibfnamefont {D.}~\bibnamefont
		{Gerace}}, \bibinfo {author} {\bibfnamefont {J.-M.}\ \bibnamefont
		{G{\'e}rard}}, \bibinfo {author} {\bibfnamefont {M.~F.}~\bibnamefont
		{{Santos}}}, \bibinfo {author} {\bibfnamefont {L.~C.}\
		\bibnamefont {Andreani}}, \ and\ \bibinfo {author} {\bibfnamefont {J.-P.}\
		\bibnamefont {Poizat}},\ }\bibfield  {title} {\enquote {\bibinfo {title}
		{Controlling the dynamics of a coupled atom-cavity system by pure
			dephasing},}\ }\href@noop {} {\bibfield  {journal} {\bibinfo  {journal}
		{Phys. Rev. B}\ }\textbf {\bibinfo {volume} {81}},\ \bibinfo {pages} {245419}
	(\bibinfo {year} {2010})}\BibitemShut {NoStop}%
\bibitem [{\citenamefont {Meldrum}\ \emph {et~al.}(2010)\citenamefont
	{Meldrum}, \citenamefont {Bianucci},\ and\ \citenamefont
	{Marsiglio}}]{Meldrum10}%
\BibitemOpen
\bibfield  {author} {\bibinfo {author} {\bibfnamefont {A.}~\bibnamefont
		{Meldrum}}, \bibinfo {author} {\bibfnamefont {P.}~\bibnamefont {Bianucci}}, \
	and\ \bibinfo {author} {\bibfnamefont {F.}~\bibnamefont {Marsiglio}},\
}\bibfield  {title} {\enquote {\bibinfo {title} {Modification of ensemble
		emission rates and luminescence spectra for inhomogeneously broadened
		distributions of quantum dots coupled to optical microcavities},}\
}\href@noop {} {\bibfield  {journal} {\bibinfo  {journal} {Optics Express}\
}\textbf {\bibinfo {volume} {18}},\ \bibinfo {pages} {10230} (\bibinfo {year}
{2010})}\BibitemShut {NoStop}%
\bibitem [{\citenamefont {Gali}\ and\ \citenamefont {Maze}(2013)}]{Gali13}%
\BibitemOpen
\bibfield  {author} {\bibinfo {author} {\bibfnamefont {Adam}\ \bibnamefont
		{Gali}}\ and\ \bibinfo {author} {\bibfnamefont {Jeronimo~R.}\ \bibnamefont
		{Maze}},\ }\bibfield  {title} {\enquote {\bibinfo {title} {\textit{Ab initio}
			study of the split silicon-vacancy defect in diamond: Electronic structure
			and related properties},}\ }\href@noop {} {\bibfield  {journal} {\bibinfo
		{journal} {Phys. Rev. B}\ }\textbf {\bibinfo {volume} {88}},\ \bibinfo
	{pages} {235205} (\bibinfo {year} {2013})}\BibitemShut {NoStop}%
\bibitem [{\citenamefont {Wang}\ \emph {et~al.}(2006)\citenamefont {Wang},
	\citenamefont {Kurtsiefer}, \citenamefont {Weinfurter},\ and\ \citenamefont
	{Burchard}}]{Wang06}%
\BibitemOpen
\bibfield  {author} {\bibinfo {author} {\bibfnamefont {Chunlang}\
		\bibnamefont {Wang}}, \bibinfo {author} {\bibfnamefont {Christian}\
		\bibnamefont {Kurtsiefer}}, \bibinfo {author} {\bibfnamefont {Harald}\
		\bibnamefont {Weinfurter}}, \ and\ \bibinfo {author} {\bibfnamefont {Bernd}\
		\bibnamefont {Burchard}},\ }\bibfield  {title} {\enquote {\bibinfo {title}
		{Single photon emission from siv centres in diamond produced by ion
			implantation},}\ }\href@noop {} {\bibfield  {journal} {\bibinfo  {journal}
		{Journal of Physics B: Atomic, Molecular and Optical Physics}\ }\textbf
	{\bibinfo {volume} {39}},\ \bibinfo {pages} {37} (\bibinfo {year}
	{2006})}\BibitemShut {NoStop}%
\bibitem [{\citenamefont {Neu}\ \emph {et~al.}(2011{\natexlab{a}})\citenamefont
	{Neu}, \citenamefont {Steinmetz}, \citenamefont {Riedrich-M\"oller},
	\citenamefont {Gsell}, \citenamefont {Fischer}, \citenamefont {Schreck},\
	and\ \citenamefont {Becher}}]{Neu11}%
\BibitemOpen
\bibfield  {author} {\bibinfo {author} {\bibfnamefont {Elke}\ \bibnamefont
		{Neu}}, \bibinfo {author} {\bibfnamefont {David}\ \bibnamefont {Steinmetz}},
	\bibinfo {author} {\bibfnamefont {Janine}\ \bibnamefont {Riedrich-M\"oller}},
	\bibinfo {author} {\bibfnamefont {Stefan}\ \bibnamefont {Gsell}}, \bibinfo
	{author} {\bibfnamefont {Martin}\ \bibnamefont {Fischer}}, \bibinfo {author}
	{\bibfnamefont {Matthias}\ \bibnamefont {Schreck}}, \ and\ \bibinfo {author}
	{\bibfnamefont {Christoph}\ \bibnamefont {Becher}},\ }\bibfield  {title}
{\enquote {\bibinfo {title} {Single photon emission from silicon-vacancy
			colour centres in chemical vapour deposition nano-diamonds on iridium},}\
}\href@noop {} {\bibfield  {journal} {\bibinfo  {journal} {New Journal of
		Physics}\ }\textbf {\bibinfo {volume} {13}},\ \bibinfo {pages} {025012}
(\bibinfo {year} {2011}{\natexlab{a}})}\BibitemShut {NoStop}%
\bibitem [{\citenamefont {Neu}\ \emph {et~al.}(2012)\citenamefont {Neu},
	\citenamefont {Agio},\ and\ \citenamefont {Becher}}]{Neu12}%
\BibitemOpen
\bibfield  {author} {\bibinfo {author} {\bibfnamefont {Elke}\ \bibnamefont
		{Neu}}, \bibinfo {author} {\bibfnamefont {Mario}\ \bibnamefont {Agio}}, \
	and\ \bibinfo {author} {\bibfnamefont {Christoph}\ \bibnamefont {Becher}},\
}\bibfield  {title} {\enquote {\bibinfo {title} {Photophysics of single
		silicon vacancy centers in diamond: implications for single photon
		emission},}\ }\href@noop {} {\bibfield  {journal} {\bibinfo  {journal}
	{Optics Express}\ }\textbf {\bibinfo {volume} {20}},\ \bibinfo {pages}
{19956} (\bibinfo {year} {2012})}\BibitemShut {NoStop}%
\bibitem [{\citenamefont {Rogers}\ \emph {et~al.}(2014)\citenamefont {Rogers},
	\citenamefont {Jahnke}, \citenamefont {Teraji}, \citenamefont {Marseglia},
	\citenamefont {M{\"u}ller}, \citenamefont {Naydenov}, \citenamefont
	{Schauffert}, \citenamefont {Kranz}, \citenamefont {Isoya}, \citenamefont
	{McGuinnes},\ and\ \citenamefont {Jelezko}}]{Rogers14}%
\BibitemOpen
\bibfield  {author} {\bibinfo {author} {\bibfnamefont {L.~J.}\ \bibnamefont
		{Rogers}}, \bibinfo {author} {\bibfnamefont {K.~D.}\ \bibnamefont {Jahnke}},
	\bibinfo {author} {\bibfnamefont {T.}~\bibnamefont {Teraji}}, \bibinfo
	{author} {\bibfnamefont {L.}~\bibnamefont {Marseglia}}, \bibinfo {author}
	{\bibfnamefont {C.}~\bibnamefont {M{\"u}ller}}, \bibinfo {author}
	{\bibfnamefont {B.}~\bibnamefont {Naydenov}}, \bibinfo {author}
	{\bibfnamefont {H.}~\bibnamefont {Schauffert}}, \bibinfo {author}
	{\bibfnamefont {C.}~\bibnamefont {Kranz}}, \bibinfo {author} {\bibfnamefont
		{J.}~\bibnamefont {Isoya}}, \bibinfo {author} {\bibfnamefont {L.~P.}\
		\bibnamefont {McGuinnes}}, \ and\ \bibinfo {author} {\bibfnamefont
		{F.}~\bibnamefont {Jelezko}},\ }\bibfield  {title} {\enquote {\bibinfo
		{title} {Multiple intrinsically identical single-photon emitters in the solid
			state},}\ }\href@noop {} {\bibfield  {journal} {\bibinfo  {journal} {Nature
			Communications}\ }\textbf {\bibinfo {volume} {5}},\ \bibinfo {pages} {4739}
	(\bibinfo {year} {2014})}\BibitemShut {NoStop}%
\bibitem [{\citenamefont {Hunger}\ \emph {et~al.}(2010)\citenamefont {Hunger},
	\citenamefont {Steinmetz}, \citenamefont {Colombe}, \citenamefont {Deutsch},
	\citenamefont {H{\"a}nsch},\ and\ \citenamefont {Reichel}}]{Hunger10b}%
\BibitemOpen
\bibfield  {author} {\bibinfo {author} {\bibfnamefont {D.}~\bibnamefont
		{Hunger}}, \bibinfo {author} {\bibfnamefont {T.}~\bibnamefont {Steinmetz}},
	\bibinfo {author} {\bibfnamefont {Y.}~\bibnamefont {Colombe}}, \bibinfo
	{author} {\bibfnamefont {C.}~\bibnamefont {Deutsch}}, \bibinfo {author}
	{\bibfnamefont {T.~W.}\ \bibnamefont {H{\"a}nsch}}, \ and\ \bibinfo {author}
	{\bibfnamefont {J.}~\bibnamefont {Reichel}},\ }\bibfield  {title} {\enquote
	{\bibinfo {title} {Fiber fabry-perot cavity with high finesse},}\ }\href@noop
{} {\bibfield  {journal} {\bibinfo  {journal} {New J. Phys.}\ }\textbf
	{\bibinfo {volume} {12}},\ \bibinfo {pages} {065038} (\bibinfo {year}
	{2010})}\BibitemShut {NoStop}%
\bibitem [{\citenamefont {Hunger}\ \emph {et~al.}(2012)\citenamefont {Hunger},
	\citenamefont {Deutsch}, \citenamefont {Barbour}, \citenamefont {Warburton},\
	and\ \citenamefont {Reichel}}]{Hunger12}%
\BibitemOpen
\bibfield  {author} {\bibinfo {author} {\bibfnamefont {D.}~\bibnamefont
		{Hunger}}, \bibinfo {author} {\bibfnamefont {C.}~\bibnamefont {Deutsch}},
	\bibinfo {author} {\bibfnamefont {R.~J.}\ \bibnamefont {Barbour}}, \bibinfo
	{author} {\bibfnamefont {R.~J.}\ \bibnamefont {Warburton}}, \ and\ \bibinfo
	{author} {\bibfnamefont {J.}~\bibnamefont {Reichel}},\ }\bibfield  {title}
{\enquote {\bibinfo {title} {Laser micro-fabrication of concave,
			low-roughness features in silica},}\ }\href@noop {} {\bibfield  {journal}
	{\bibinfo  {journal} {AIP Advances}\ }\textbf {\bibinfo {volume} {2}},\
	\bibinfo {pages} {012119} (\bibinfo {year} {2012})}\BibitemShut {NoStop}%
\bibitem [{\citenamefont {Benedikter}\ \emph {et~al.}(2015)\citenamefont
	{Benedikter}, \citenamefont {H{\"u}mmer}, \citenamefont {Mader},
	\citenamefont {Schlederer}, \citenamefont {Reichel}, \citenamefont
	{H{\"a}nsch},\ and\ \citenamefont {Hunger}}]{Benedikter15}%
\BibitemOpen
\bibfield  {author} {\bibinfo {author} {\bibfnamefont {J.}~\bibnamefont
		{Benedikter}}, \bibinfo {author} {\bibfnamefont {T.}~\bibnamefont
		{H{\"u}mmer}}, \bibinfo {author} {\bibfnamefont {M.}~\bibnamefont {Mader}},
	\bibinfo {author} {\bibfnamefont {B.}~\bibnamefont {Schlederer}}, \bibinfo
	{author} {\bibfnamefont {J.}~\bibnamefont {Reichel}}, \bibinfo {author}
	{\bibfnamefont {T.W.}\ \bibnamefont {H{\"a}nsch}}, \ and\ \bibinfo {author}
	{\bibfnamefont {D.}~\bibnamefont {Hunger}},\ }\bibfield  {title} {\enquote
	{\bibinfo {title} {Transverse-mode mixing and diffraction loss in tunable
			fabry-perot microcavities},}\ }\href@noop {} {\bibfield  {journal} {\bibinfo
		{journal} {New J. Phys.}\ }\textbf {\bibinfo {volume} {17}},\ \bibinfo
	{pages} {053051} (\bibinfo {year} {2015})}\BibitemShut {NoStop}%
\bibitem [{\citenamefont {Neu}\ \emph {et~al.}(2011{\natexlab{b}})\citenamefont
	{Neu}, \citenamefont {Arend}, \citenamefont {Gross}, \citenamefont {Guldner},
	\citenamefont {Hepp}, \citenamefont {Steinmetz}, \citenamefont {Zscherpel},
	\citenamefont {Ghodbane}, \citenamefont {Sternschulte}, \citenamefont
	{Steinm\"uller-Nethl}, \citenamefont {Liang}, \citenamefont {Krueger},\ and\
	\citenamefont {Becher}}]{Neu11b}%
\BibitemOpen
\bibfield  {author} {\bibinfo {author} {\bibfnamefont {E.}~\bibnamefont
		{Neu}}, \bibinfo {author} {\bibfnamefont {C.}~\bibnamefont {Arend}}, \bibinfo
	{author} {\bibfnamefont {E.}~\bibnamefont {Gross}}, \bibinfo {author}
	{\bibfnamefont {F.}~\bibnamefont {Guldner}}, \bibinfo {author} {\bibfnamefont
		{C.}~\bibnamefont {Hepp}}, \bibinfo {author} {\bibfnamefont {D.}~\bibnamefont
		{Steinmetz}}, \bibinfo {author} {\bibfnamefont {E.}~\bibnamefont
		{Zscherpel}}, \bibinfo {author} {\bibfnamefont {S.}~\bibnamefont {Ghodbane}},
	\bibinfo {author} {\bibfnamefont {H.}~\bibnamefont {Sternschulte}}, \bibinfo
	{author} {\bibfnamefont {D.}~\bibnamefont {Steinm\"uller-Nethl}}, \bibinfo
	{author} {\bibfnamefont {Y.}~\bibnamefont {Liang}}, \bibinfo {author}
	{\bibfnamefont {A.}~\bibnamefont {Krueger}}, \ and\ \bibinfo {author}
	{\bibfnamefont {C.}~\bibnamefont {Becher}},\ }\bibfield  {title} {\enquote
	{\bibinfo {title} {Narrowband fluorescent nanodiamonds produced from chemical
			vapor deposition films},}\ }\href@noop {} {\bibfield  {journal} {\bibinfo
		{journal} {Applied Physics Letters}\ }\textbf {\bibinfo {volume} {98}}
	(\bibinfo {year} {2011}{\natexlab{b}})}\BibitemShut {NoStop}%
\bibitem [{\citenamefont {Mader}\ \emph {et~al.}(2015)\citenamefont {Mader},
	\citenamefont {Reichel}, \citenamefont {H{\"a}nsch},\ and\ \citenamefont
	{Hunger}}]{Mader15}%
\BibitemOpen
\bibfield  {author} {\bibinfo {author} {\bibfnamefont {M.}~\bibnamefont
		{Mader}}, \bibinfo {author} {\bibfnamefont {J.}~\bibnamefont {Reichel}},
	\bibinfo {author} {\bibfnamefont {T.~W.}\ \bibnamefont {H{\"a}nsch}}, \ and\
	\bibinfo {author} {\bibfnamefont {D.}~\bibnamefont {Hunger}},\ }\bibfield
{title} {\enquote {\bibinfo {title} {A scanning cavity microscope},}\
}\href@noop {} {\bibfield  {journal} {\bibinfo  {journal} {Nature
		Communications}\ }\textbf {\bibinfo {volume} {6}},\ \bibinfo {pages} {7249}
(\bibinfo {year} {2015})}\BibitemShut {NoStop}%
\bibitem [{\citenamefont {Arend}\ \emph {et~al.}(2016)\citenamefont {Arend},
	\citenamefont {Becker}, \citenamefont {Sternschulte}, \citenamefont
	{Steinm\"uller-Nethl},\ and\ \citenamefont {Becher}}]{Arend16}%
\BibitemOpen
\bibfield  {author} {\bibinfo {author} {\bibfnamefont {Carsten}\ \bibnamefont
		{Arend}}, \bibinfo {author} {\bibfnamefont {Jonas~Nils}\ \bibnamefont
		{Becker}}, \bibinfo {author} {\bibfnamefont {Hadwig}\ \bibnamefont
		{Sternschulte}}, \bibinfo {author} {\bibfnamefont {Doris}\ \bibnamefont
		{Steinm\"uller-Nethl}}, \ and\ \bibinfo {author} {\bibfnamefont {Christoph}\
		\bibnamefont {Becher}},\ }\bibfield  {title} {\enquote {\bibinfo {title}
		{Photoluminescence excitation and spectral hole burning spectroscopy of
			silicon vacancy centers in diamond},}\ }\href@noop {} {\bibfield  {journal}
	{\bibinfo  {journal} {Phys. Rev. B}\ }\textbf {\bibinfo {volume} {94}},\
	\bibinfo {pages} {045203} (\bibinfo {year} {2016})}\BibitemShut {NoStop}%
\bibitem [{\citenamefont {Feng}\ and\ \citenamefont {Schwartz}(1993)}]{Feng93}%
\BibitemOpen
\bibfield  {author} {\bibinfo {author} {\bibfnamefont {T.}~\bibnamefont
		{Feng}}\ and\ \bibinfo {author} {\bibfnamefont {B.~D.}\ \bibnamefont
		{Schwartz}},\ }\bibfield  {title} {\enquote {\bibinfo {title}
		{Characteristics and origin of the 1.681 ev luminescence center in
			chemical-vapor-deposited diamond films},}\ }\href@noop {} {\bibfield
	{journal} {\bibinfo  {journal} {J. Appl. Phys.}\ }\textbf {\bibinfo {volume}
		{73}},\ \bibinfo {pages} {1415} (\bibinfo {year} {1993})}\BibitemShut
{NoStop}%
\bibitem [{\citenamefont {Grange}\ \emph {et~al.}(2015)\citenamefont {Grange},
	\citenamefont {Hornecker}, \citenamefont {Hunger}, \citenamefont {Poizat},
	\citenamefont {G{\'e}rard}, \citenamefont {Senellart},\ and\ \citenamefont
	{Auff{\`e}ves}}]{Grange15}%
\BibitemOpen
\bibfield  {author} {\bibinfo {author} {\bibfnamefont {Thomas}\ \bibnamefont
		{Grange}}, \bibinfo {author} {\bibfnamefont {Gaston}\ \bibnamefont
		{Hornecker}}, \bibinfo {author} {\bibfnamefont {David}\ \bibnamefont
		{Hunger}}, \bibinfo {author} {\bibfnamefont {Jean-Philippe}\ \bibnamefont
		{Poizat}}, \bibinfo {author} {\bibfnamefont {Jean-Michel}\ \bibnamefont
		{G{\'e}rard}}, \bibinfo {author} {\bibfnamefont {Pascale}\ \bibnamefont
		{Senellart}}, \ and\ \bibinfo {author} {\bibfnamefont {Alexia}\ \bibnamefont
		{Auff{\`e}ves}},\ }\bibfield  {title} {\enquote {\bibinfo {title}
		{Cavity-funneled generation of indistinguishable single photons from strongly
			dissipative quantum emitters},}\ }\href@noop {} {\bibfield  {journal}
	{\bibinfo  {journal} {Physical review letters}\ }\textbf {\bibinfo {volume}
		{114}},\ \bibinfo {pages} {193601} (\bibinfo {year} {2015})}\BibitemShut
{NoStop}%
\bibitem [{\citenamefont {Knill}\ \emph {et~al.}(2007)\citenamefont {Knill},
	\citenamefont {Laflamme},\ and\ \citenamefont {Milburn}}]{Knill00}%
\BibitemOpen
\bibfield  {author} {\bibinfo {author} {\bibfnamefont {E.}~\bibnamefont
		{Knill}}, \bibinfo {author} {\bibfnamefont {R.}~\bibnamefont {Laflamme}}, \
	and\ \bibinfo {author} {\bibfnamefont {G.~J.}\ \bibnamefont {Milburn}},\
}\bibfield  {title} {\enquote {\bibinfo {title} {A scheme for efficient
		quantum computation with linear optics},}\ }\href@noop {} {\bibfield
{journal} {\bibinfo  {journal} {Nature}\ }\textbf {\bibinfo {volume} {409}},\
\bibinfo {pages} {46--52} (\bibinfo {year} {2007})}\BibitemShut {NoStop}%
\bibitem [{\citenamefont {Kok}\ \emph {et~al.}(2007)\citenamefont {Kok},
	\citenamefont {Munro}, \citenamefont {Nemoto}, \citenamefont {Ralph},
	\citenamefont {Dowling},\ and\ \citenamefont {Milburn}}]{Kok07}%
\BibitemOpen
\bibfield  {author} {\bibinfo {author} {\bibfnamefont {Pieter}\ \bibnamefont
		{Kok}}, \bibinfo {author} {\bibfnamefont {W.~J.}\ \bibnamefont {Munro}},
	\bibinfo {author} {\bibfnamefont {Kae}\ \bibnamefont {Nemoto}}, \bibinfo
	{author} {\bibfnamefont {T.~C.}\ \bibnamefont {Ralph}}, \bibinfo {author}
	{\bibfnamefont {Jonathan~P.}\ \bibnamefont {Dowling}}, \ and\ \bibinfo
	{author} {\bibfnamefont {G.~J.}\ \bibnamefont {Milburn}},\ }\bibfield
{title} {\enquote {\bibinfo {title} {Linear optical quantum computing with
			photonic qubits},}\ }\href@noop {} {\bibfield  {journal} {\bibinfo  {journal}
		{Rev. Mod. Phys.}\ }\textbf {\bibinfo {volume} {79}},\ \bibinfo {pages}
	{135--174} (\bibinfo {year} {2007})}\BibitemShut {NoStop}%
\bibitem [{\citenamefont {Furman}\ and\ \citenamefont
	{Tikhonravov}(1992)}]{Furman92}%
\BibitemOpen
\bibfield  {author} {\bibinfo {author} {\bibfnamefont {S.~A.}\ \bibnamefont
		{Furman}}\ and\ \bibinfo {author} {\bibfnamefont {A.~V.}\ \bibnamefont
		{Tikhonravov}},\ }\href@noop {} {\emph {\bibinfo {title} {Basics of Optics of
			Multilayer Systems}}}\ (\bibinfo  {publisher} {World Scientific Publishing},\
\bibinfo {year} {1992})\BibitemShut {NoStop}%
\bibitem [{\citenamefont {Hood}\ \emph {et~al.}(2001)\citenamefont {Hood},
	\citenamefont {Kimble},\ and\ \citenamefont {Ye}}]{Hood01}%
\BibitemOpen
\bibfield  {author} {\bibinfo {author} {\bibfnamefont {C.~J.}\ \bibnamefont
		{Hood}}, \bibinfo {author} {\bibfnamefont {H.~J.}\ \bibnamefont {Kimble}}, \
	and\ \bibinfo {author} {\bibfnamefont {J.}~\bibnamefont {Ye}},\ }\bibfield
{title} {\enquote {\bibinfo {title} {Characterization of high finesse
			mirrors: Loss, phase shifts, and mode structure in an optical cavity},}\
}\href@noop {} {\bibfield  {journal} {\bibinfo  {journal} {Phys. Rev. A}\
}\textbf {\bibinfo {volume} {64}},\ \bibinfo {pages} {033804} (\bibinfo
{year} {2001})}\BibitemShut {NoStop}%
\bibitem [{\citenamefont {Lukosz}(1979)}]{Lukosz79}%
\BibitemOpen
\bibfield  {author} {\bibinfo {author} {\bibfnamefont {W.}~\bibnamefont
		{Lukosz}},\ }\bibfield  {title} {\enquote {\bibinfo {title} {Light emission
			by magnetic and electric dipoles close to a plane dielectric interface.
			{III}. radiation patterns of dipoles with arbitrary orientation},}\
}\href@noop {} {\ \textbf {\bibinfo {volume} {69}},\ \bibinfo {pages} {1495}
(\bibinfo {year} {1979})}\BibitemShut {NoStop}%
\bibitem [{\citenamefont {Joyce}\ and\ \citenamefont
	{{DeLoach}}(1984)}]{Joyce84}%
\BibitemOpen
\bibfield  {author} {\bibinfo {author} {\bibfnamefont {W.~B.}\ \bibnamefont
		{Joyce}}\ and\ \bibinfo {author} {\bibfnamefont {B.~C.}\ \bibnamefont
		{{DeLoach}}},\ }\bibfield  {title} {\enquote {\bibinfo {title} {Alignment of
			gaussian beams},}\ }\href@noop {} {\bibfield  {journal} {\bibinfo  {journal}
		{Applied Optics}\ }\textbf {\bibinfo {volume} {23}},\ \bibinfo {pages} {4187}
	(\bibinfo {year} {1984})}\BibitemShut {NoStop}%
\bibitem [{\citenamefont {Kleckner}\ \emph {et~al.}(2010)\citenamefont
	{Kleckner}, \citenamefont {Irvine}, \citenamefont {Oemrawsingh},\ and\
	\citenamefont {Bouwmeester}}]{Kleckner10}%
\BibitemOpen
\bibfield  {author} {\bibinfo {author} {\bibfnamefont {D.}~\bibnamefont
		{Kleckner}}, \bibinfo {author} {\bibfnamefont {W.~T.~M.}\ \bibnamefont
		{Irvine}}, \bibinfo {author} {\bibfnamefont {S.~S.~R.}\ \bibnamefont
		{Oemrawsingh}}, \ and\ \bibinfo {author} {\bibfnamefont {D.}~\bibnamefont
		{Bouwmeester}},\ }\bibfield  {title} {\enquote {\bibinfo {title}
		{Diffraction-limited high-finesse optical cavities},}\ }\href@noop {}
{\bibfield  {journal} {\bibinfo  {journal} {Phys. Rev. A}\ }\textbf {\bibinfo
		{volume} {81}},\ \bibinfo {pages} {043814} (\bibinfo {year}
	{2010})}\BibitemShut {NoStop}%
\bibitem [{\citenamefont {Li}\ \emph {et~al.}(2016)\citenamefont {Li},
	\citenamefont {Zhou}, \citenamefont {Rasmita}, \citenamefont {Aharonovich},\
	and\ \citenamefont {Gao}}]{Li16}%
\BibitemOpen
\bibfield  {author} {\bibinfo {author} {\bibfnamefont {Ke}~\bibnamefont
		{Li}}, \bibinfo {author} {\bibfnamefont {Yu}~\bibnamefont {Zhou}}, \bibinfo
	{author} {\bibfnamefont {A}~\bibnamefont {Rasmita}}, \bibinfo {author}
	{\bibfnamefont {I}~\bibnamefont {Aharonovich}}, \ and\ \bibinfo {author}
	{\bibfnamefont {WB}~\bibnamefont {Gao}},\ }\bibfield  {title} {\enquote
	{\bibinfo {title} {Nonblinking emitters with nearly lifetime-limited
			linewidths in cvd nanodiamonds},}\ }\href@noop {} {\bibfield  {journal}
	{\bibinfo  {journal} {Physical Review Applied}\ }\textbf {\bibinfo {volume}
		{6}},\ \bibinfo {pages} {024010} (\bibinfo {year} {2016})}\BibitemShut
{NoStop}%
\end{thebibliography}
\end{document}